%% file: main.tex
\documentclass[manuscript, screen]{acmart}
\usepackage{color,soul}
\usepackage[toc,page]{appendix}

\usepackage{amsmath,amssymb,amsfonts}

\usepackage[ruled,linesnumbered,algosection]{algorithm2e} 

\AtBeginDocument{%
  \providecommand\BibTeX{{%
    \normalfont B\kern-0.5em{\scshape i\kern-0.25em b}\kern-0.8em\TeX}}}

\setcopyright{none}
\renewcommand\footnotetextcopyrightpermission[1]{}
\begin{document}

\title{Collusion-Resilient Probabilistic Fingerprinting Scheme for Correlated Data}

\author{Emre Yilmaz}
\email{exy109@case.edu}
\affiliation{%
  \institution{Case Western Reserve University}
}
\author{Erman Ayday}
\email{exa208@case.edu}
\affiliation{%
  \institution{Case Western Reserve University}
}
\affiliation{%
  \institution{Bilkent University}
  \city{Ankara}
  \state{Turkey}
}
\begin{abstract}

In order to receive personalized services, individuals share their personal data (e.g., location patterns, healthcare data, or financial data) with a wide range of service providers, hoping that their data will remain confidential. 
Thus, in case of an unauthorized distribution of their personal data by these service providers (or in case of a data breach) data owners want to identify the source of such data leakage. 
Digital fingerprinting schemes have been developed to embed a hidden and unique fingerprint into shared digital content, especially multimedia, to provide such liability guarantees. However, existing techniques utilize the high redundancy in the content, which is typically not included in personal data (such as location patters or genomic data). 
In this work, we propose a probabilistic fingerprinting scheme that efficiently generates the fingerprint by considering a fingerprinting probability (to keep the data utility high) and publicly known inherent correlations between data points. 
To improve the robustness of the proposed scheme against colluding malicious service providers, we also utilize the Boneh-Shaw fingerprinting codes as a part of the proposed scheme. 
Furthermore, observing similarities between privacy-preserving data sharing techniques (that add controlled noise to the shared data) and the proposed fingerprinting scheme, we make a first attempt to develop a data sharing scheme that provides both privacy and fingerprint robustness at the same time. 
We experimentally show that fingerprint robustness and privacy have conflicting objectives and we propose a hybrid approach to control such a trade-off with a design parameter. Using the proposed hybrid approach, we show that individuals can improve their level of privacy by slightly compromising from the fingerprint robustness.
We implement and evaluate the performance of the proposed scheme on real genomic data. Our experimental results show the efficiency and robustness of the proposed scheme.
  
\end{abstract}

\keywords{Fingerprinting, Liability, Privacy, Data sharing}

\maketitle

\input{sections/introduction.tex}

\input{sections/related_work.tex}

\input{sections/model.tex}

\input{sections/algorithm.tex}

\input{sections/collusions.tex}

\input{sections/privacy.tex}

\input{sections/evaluation.tex}

\input{sections/discussion.tex}

\input{sections/conclusion.tex}

\newpage

\bibliographystyle{ACM-Reference-Format}
\bibliography{sigproc}

\end{document}

%% file: sections/introduction.tex
\vspace{6mm}
\section{Introduction} \label{sec:intro}

In today's data-driven world, individuals share vast amount of personal information with several service providers (SPs) to receive personalized services. During such data sharings, data owners usually do not want SPs to share their personal data with other third parties. Such issues are typically addressed via a consent (or data usage agreement) between a data owner and an SP to determine how much the SP gains the ownership of the user's data. However, user's data may often end up in the hands of unauthorized third parties since (i) SPs sometimes share (or sell) users' personal information without their authorization or (ii) databases of SPs are sometimes breached (e.g., due to insufficient or non-existing security measures).

When such a leakage occurs, data owners would at least like to know the source of it to keep the corresponding SP(s) liable due to the leakage. If techniques exist to help the data owners identify the source of the data leakages, SPs (knowing potential consequences) will be hesitant to share users' data without their authorization or they will take more serious security measures against data breaches. Thus, it is crucial to develop such techniques that can be used when sharing personal data with untrusted SPs and that are robust against various attacks that may be launched by malicious SPs. In this work, we propose such a technique by focusing on sequential personal information, such as genomic data, location data, or financial data.

Similar techniques exist for multimedia to prevent copyrighted content being copied or shared without the authorization of the data owner. Digital watermarking~\cite{podilchuk2001digital} is a technique to prove the ownership of digital content by embedding a mark into it, in which embedded watermarks can be the same for each copy of digital object. In order to detect the source of an illegal distribution, different mark should be used for each copy. 
Digital fingerprinting~\cite{wu2004collusion} is a technique to identify the recipient of a digital object by embedding a unique mark (called fingerprint) into the digital object. The aim of fingerprinting is to identify the guilty agent who is responsible for data leakage. Watermarking and fingerprinting techniques have been developed for different types of digital content, such as audio, video, software, relational databases, graphs, and maps. However, such techniques are not directly applicable for our scenario (sharing personal correlated data) because (i) they (especially for multimedia) utilize the high redundancy in the data, (ii) the embedded marks need to be large, which reduces the utility of shared data, and (iii) they do not consider the correlations between data points.

In this work, we propose a fingerprinting technique for sequential data having correlations between data points. We consider several different malicious behaviour that can be launched by the malicious SPs against the proposed fingerprinting scheme including: (i) flipping data points, (ii) using a subset of the data points, (iii) utilizing the correlations in the data, and (iv) colluding SPs to identify and/or distort the fingerprint. 

The proposed fingerprinting scheme essentially relies on adding controlled noise into particular data points in the original data. We build a correlation model (that captures the statistics in the data) and add the fingerprint to the data that is consistent with the nature of the data. By doing so, we avoid malicious SPs identify and distort the added fingerprint using the auxiliary information about the data model. We also consider colluding SPs (that receive different fingerprinted copies of the same data) who aim to detect and distort the fingerprints. The proposed fingerprinting scheme utilizes Boneh-Shaw codes~\cite{boneh1} to improve its collusion resistance and integrates such codes with a novel algorithm to also provide robustness against other types of malicious behavior, such as flipping or utilizing correlations in the data (that are not considered by Boneh-Shaw codes). Furthermore, we propose a detection algorithm that identifies the suspects based on the similarity of their copies with leaked data and selects one of them by checking Boneh-Shaw codes embedded in their copies. Besides providing robustness against the aforementioned attacks, the proposed scheme also keeps the data utility high by controlling the fraction of fingerprinted data points. 

In most data sharing scenarios, individuals also want to share their personal information with the SPs under certain privacy guarantees. Several existing techniques rely on adding controlled noise on the shared data to provide privacy guarantees (including large tech companies, such as Google~\cite{erlingsson2014rappor} and Apple~\cite{dp2017learning}). Realizing these similar methodologies between the proposed scheme and privacy-preserving data sharing, we also propose a scheme that provides both privacy and liability (i.e., robust fingerprinting) while sharing personal data with the SPs. Here, the main challenge is the conflicting goals between data privacy and robust fingerprinting. To provide higher privacy, the same (or similar) noise pattern should be used at each new sharing of the data (e.g., by maximizing the overlap between noisy data points across different sharings). On the other hand, for robust fingerprinting, shared noise patterns should be unique to identify the source of a potential data leakage (e.g., by minimizing the overlap between noisy data points across different sharings). Thus, we propose a scheme that controls the size of this overlapping region to identify a sweet-spot between privacy and robust fingerprinting.

We implement the proposed fingerprinting scheme for genomic data sharing using a real-life genomic dataset. Via simulations, we show the robustness of the proposed scheme against various types of attacks that can be launched against a fingerprinting scheme. We show that the proposed scheme is efficient and scalable in terms of its running time. We also study the balance between fingerprint robustness and data privacy via simulations on genomic data.

The rest of the paper is organized as follows. We summarize the related work in Section \ref{sec:related_work}. We describe the problem settings in Section \ref{sec:model}. We propose the probabilistic fingerprinting scheme in Section \ref{sec:scheme} and explain how to utilize Boneh-Shaw codes to resist collusion attacks in Section \ref{sec:collusion}. We propose a hybrid approach in Section \ref{sec:privacy} to provide privacy and liability together. We present our experimental results in Section \ref{sec:eval}. We discuss more about our scheme based on our evaluation results in Section \ref{sec:discuss} and conclude the paper in Section \ref{sec:conc}.

%% file: sections/related_work.tex
\section{Related Work} 
\label{sec:related_work}

Digital watermarking is the act of embedding an owner-specific mark into a digital object (e.g., an image, song, or video) to prove the ownership of the object~\cite{cox2002digital}. Digital watermarks are typically used for copyright and copy protection of multimedia content~\cite{ buyer-seller, chung1998digital}. The techniques to watermark audio ~\cite{ bassia2001robust, ko2005time}, image ~\cite{Chang, Wang2}, and video ~\cite{swanson1998multiresolution,  hartung1998watermarking} have been developed. Watermarking techniques have also been proposed for text documents \cite{Brassil1999,Atallah2,Topkara1}, graphs \cite{qu1999hiding, gross2011query, zhao2015towards}, maps \cite{yan2011key, wang2012watermarking, lee2013vector}, time-series data such as electrocardiograms \cite{kozat2009embedding, soltani2014lightweight}, and spatiotemporal (trajectory) datasets \cite{jin2005watermarking, lucchese2010rights}. In general, watermarking schemes (i) benefit from the high redundancy in the data, (ii) mostly aim to prove ownership of the digital object, and (ii) do not consider robustness against various types of attacks (that are discussed in Section~\ref{sec:attack}). 

Digital fingerprinting is similar to watermarking in a sense that it also embeds a mark into the object. However, fingerprinting can be seen as a personalized version of watermarking since embedded mark (i.e., fingerprint) is different in each copy of data with the objective to identify the recipient if data is disclosed to a third party without authorization of data owner. 
Since each copy is different in fingerprinting, malicious recipients can collude and detect fingerprinted points by comparing their copies. Boneh and Shaw proposed a general fingerprinting solution for binary data that is robust against collusion~\cite{boneh1}. Their scheme constructs fingerprinting codes in such a way that the attackers cannot detect some of the fingerprints due to inclusion of overlapped fingerprints in each copy. However, fingerprint length needs to be significantly long to guarantee robustness against collusion, which reduces the utility of the data. Furthermore, they do not consider complex attacks against the fingerprinting algorithms, such as using correlations in the data for detecting fingerprinted points. Due to such limitations, their codes cannot be used directly for personal data sharing. As we explain in Section~\ref{sec:use-boneh-shaw}, we utilize Boneh-Shaw codes to improve the robustness of the proposed scheme against collusion attacks.

Some other fingerprinting schemes have been proposed for multimedia~\cite{wu2004collusion}, relational databases~\cite{li2005fingerprinting, liu2004block, lafaye2008watermill}, and sequential data~\cite{ayday2019robust}. Similar to multimedia watermarking schemes, fingerprinting schemes for multimedia also utilize the high redundancy in digital object, which are not applicable to personal data sharing (which typically includes less redundancy). Fingerprinting schemes for relational databases~\cite{li2005fingerprinting, liu2004block, lafaye2008watermill} insert fingerprint by altering the least significant bits of some attributes in selected tuples. Since databases consist of numerous tuples, the redundancy is much higher compared to personal data. Moreover, these techniques do not consider correlations between attributes. The scheme proposed for sequential data~\cite{ayday2019robust} is inefficient due to solving an optimization problem in each step of the algorithm. In addition, the objective of the optimization problem (i.e., minimizing the probability of identifying the whole fingerprint in a collusion attack) is unrealistic, because the attackers do not need to identify the whole fingerprint to perform a successful attack; they can achieve their goal by modifying some of the fingerprinted points. Therefore, in this work we develop an efficient fingerprinting scheme which is robust against various types of attacks. Furthermore, to the best of our knowledge, for the first time we study the fingerprinting robustness and privacy together in a data sharing scheme.

%% file: sections/model.tex
\section{System Model, Threat Model, and Robustness Measures}
\label{sec:model}

In this section, we present our system model, threat model, and robustness measures.

\subsection{System and Data Model}\label{sec:system}

\begin{figure}
\centering
\includegraphics[width=14cm,keepaspectratio]{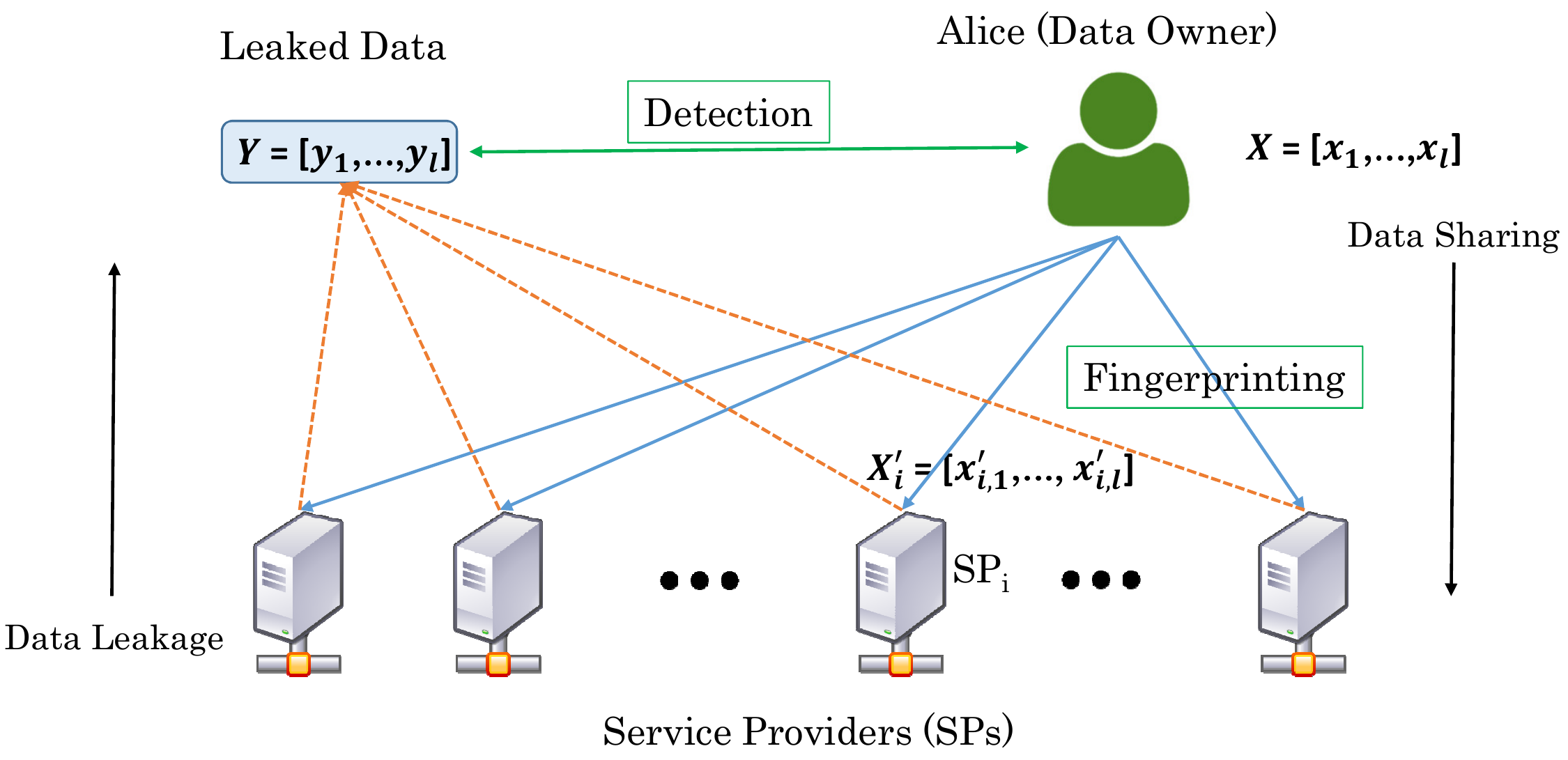}
\caption{The system model. Alice shares her data with several SPs after fingerprinting. If Alice observes that her data is leaked, her goal is to identify the guilty SP (via the detection algorithm) that is responsible for this leakage.}
\label{fig:systemModel}
\end{figure}

We show our system model in Figure~\ref{fig:systemModel}.
We assume a data owner (Alice) with a sequence of data points $\mathcal{X} = \lbrack x_1, x_2, \ldots, x_l \rbrack$, where $l$ is the length of the data and each $x_i$ can have a value from the set $\mathcal{D} = \{d_1, d_2, \ldots, d_m\}$. Alice wants to share her data with multiple service providers (SPs) to receive a service from these SPs related to her data. We represent these SPs with a set $\mathcal{S}$ and each SP with an index such that $SP_i \in \mathcal{S}$. We assume Alice wants to add a unique fingerprint in each sharing to detect the source of data leakage in the case of an unauthorized data sharing by any of these SPs. Hence, for each $SP_i \in \mathcal{S}$, Alice creates a unique fingerprinted copy $\mathcal{X}_i' = \lbrack x'_{i,1}, \ldots,x'_{i,l} \rbrack$ of original data $\mathcal{X}$ by changing the values of some data points. We list the frequently used symbols in Table~\ref{table:notationTable}.

Changing the states of more data points for fingerprinting increases the chance of Alice to detect the malicious SP(s) who leaks her data. However, fingerprinting naturally degrades the utility of shared data and one of our goals is to minimize this degradation while providing a robust fingerprinting scheme. We define the utility of a shared copy $\mathcal{X}_i'$ as $\mathcal{U}_i = 
(\sum_{j = 1}^{l}{u_j D_j^i})/(\sum_{j = 1}^{l}{u_j})$, where $u_j$ is the utility of data point $x_j$, $D_j^i = 1$ if $x_j = x'_{i,j}$ (i.e., correctly shared), and $D_j^i = -1$ if $x_j \neq x'_{i,j}$ (i.e., fingerprinted). 

On one hand, the aim of Alice is to detect the source of data leakage in case her data is shared without her consent by an SP (or collusion of multiple SPs). On the other hand, the aim of the malicious SPs is to avoid being detected while sharing as many data points as possible. A malicious $SP_i$ may avoid being detected by Alice by changing the states of some data points in its copy $\mathcal{X}_i'$ or excluding them in its unauthorized sharing. Multiple malicious SPs may also collude to detect fingerprinted data points. Furthermore, background knowledge about the data can be used to detect fingerprinted data points (we discuss more on the threat model in Section~\ref{sec:attack}). One such auxiliary information that malicious SPs can use is correlations in the data.

We assume that data is correlated and in this work, for clarity of presentation, we consider pairwise correlations between consecutive data points. The proposed mechanism can also be extended to consider more complex correlations, which may result in eliminating more possible values with low correlations. Thus, we let $P(x_{j+1} = d_{\beta} | x_j = d_{\alpha})$ values be publicly available for any $j \in \{1, \ldots,l-1\}$ and $d_{\alpha}, d_{\beta} \in \mathcal{D}$. This model represents the inherent characteristics of different data types, such as location patterns or genomic data. For instance, consecutive data points that are collected with small differences in time are correlated in location patterns. Similarly, in genomic data, point mutations (e.g., single nucleotide polymorphisms) may have pairwise correlations between each other (e.g., linkage disequilibrium~\cite{Slatkin2008}). Therefore, a malicious SP may use such correlations to detect and exclude fingerprinted data points (e.g., the ones that are not compliant with the expected correlation model) in its unauthorized data sharing.
  
\begin{table}
\small
		\centering
		\begin{tabular}{|p{0.158\columnwidth}|p{0.792\columnwidth}|}
			\hline
			$\mathcal{X} = \lbrack x_1, \ldots, x_l \rbrack$ & Original data owned by Alice (data owner). \\ \hline
			$\mathcal{D} = \{d_1, \ldots, d_m\}$ & Possible values (states) of a data point. \\ \hline
			$\mathcal{X}_i' = \lbrack x'_{i,1}, \ldots,x'_{i,l} \rbrack$ & Fingerprinted copy shared with $SP_i$. \\ \hline
			$\mathcal{Y} = \lbrack y_1, y_2, \ldots, y_l \rbrack$ & Data leaked by malicious SP(s). \\ \hline
			$\mathcal{S}$ & The set of all SPs receiving a fingerprinted copy from Alice. \\ \hline
			$\mathcal{C}$ & The set of colluding SPs in the collusion attack. \\ \hline
			$P_{x_j,d_k}$ & The probability of sharing data point $x_j$ as $d_k$ in the proposed scheme. \\ \hline
			$n$ & The number of colluding SPs in the collusion attack. \\ \hline
			$p$ & The probability of fingerprinting a data point in the proposed scheme. \\ \hline
			$p_f$, $p_s$ & The probability of flipping a data point in the flipping attack and removing a data point in the subset attack. \\ \hline
			$p_e$ & The estimation of $p$ by the colluding SPs. If $p$ is publicly known, $p_e=p$.\\ \hline
		\end{tabular}
		\caption{Frequently used symbols and notations.}
		\vspace{-5mm}
		\label{table:notationTable}
	\end{table}

\subsection{Threat Model}\label{sec:attack}
In this section, we present the attacks which may be performed by malicious SPs (attackers). We consider these attacks when developing the proposed scheme and we evaluate the robustness of the proposed scheme against these attacks in Section~\ref{sec:eval}. In all of these attacks, the main goals of the attacker(s) are (i) to avoid being detected by the data owner and (ii) to share as much correct data point as possible (i.e., not to further distort the data). The attacker(s) needs to modify the values of some data points in its copy to distort the fingerprint. These modifications in the data mostly cause utility loss for the attacker(s). Note however that in some cases, such as in the collusion attack, utility of the attackers may improve as a result of the attack. Let $\mathcal{Y} = \lbrack y_1, \ldots, y_l \rbrack$ be the leaked copy of Alice's data $\mathcal{X}$. The utility of the attacker(s) is defined as 
$\mathcal{U}_\mathcal{Y} = (\sum_{j =1}^{l}{u_j D_j}) / (\sum_{j = 1}^{l}{u_j})$,
where $u_j$ is the utility of $x_j$, $D_j = 1$ if $x_j = y_{j}$ and $D_j = -1$ if $x_j \neq y_{j}$. 

\subsubsection{Flipping Attack}\label{sec:threat:flipping}

In this attack, a malicious SP flips the values of some data points randomly to distort the fingerprint (before it does the unauthorized sharing). The malicious SP flips each data point with probability $p_f$. If it decides to flip, then it selects one of the remaining $(m-1)$ values (states) of the corresponding data point with equal probability and shares that state. Higher $p_f$ decreases the probability of being detected by Alice. However, the utility of shared data becomes lower for higher $p_f$ values. This attack does not change the size of data.

\subsubsection{Subset Attack}\label{sec:threat:subset}

This attack is similar to flipping attack, but here, a malicious SP excludes (removes) some randomly chosen data points before leaking data, instead of flipping them. We denote the probability of excluding a data point as $p_s$. This attack is not as powerful as the flipping attack because flipping data points might create a fingerprint pattern that looks similar to some other SP's fingerprint pattern, and hence Alice may falsely accuse an innocent SP. However, to succeed in the subset attack, the malicious SP needs to exclude almost all of the fingerprinted data points.

\subsubsection{Correlation Attack}\label{sec:threat:correlation}

As discussed, the correlations between consecutive data points are assumed to be publicly known. A malicious SP can use these correlations to degrade the robustness of the fingerprinting mechanism (we will define the robustness of the mechanism later). Assume the malicious $SP_i$ receives $\mathcal{X}_i' = \{x'_{i,1}, \ldots,x'_{i,l}\}$ from Alice and let two data points in the received data be $x'_{i,j+1} = d_{\beta}$ and $x'_{i,j} = d_{\alpha}$. If $P(x_{j+1} = d_{\beta} | x_j = d_{\alpha})$ is low, then the attacker infers that either $x'_{i,j}$ or $x'_{i,j+1}$ is fingerprinted with a high probability. After detecting such a pair, the malicious SP may change their values or exclude them before unauthorized sharing. In other words, the malicious SP combines the correlation attack with flipping or subset attacks. 

Since flipping attack can be more powerful (as discussed in Section~\ref{sec:threat:subset}), we assume the malicious SP combines correlation attack with flipping attack as follows: For two consecutive data points $x_j$ and $x_{j+1}$, the malicious $SP_i$ checks the conditional probability of $x_{j+1}$ having the value of $x'_{i,j+1}$ given $x_j$ having the value of $x'_{i,j}$. If this probability is less than a threshold $\tau_c$, the malicious SP changes the value of $x'_{i,j+1}$ to a different value from the set $\mathcal{D}$ that provides the highest conditional probability (correlation) between $x'_{i,j}$ and $x'_{i,j+1}$. Otherwise (i.e., if the the conditional probability is greater than or equal to $\tau_c$), the malicious SP flips the value of $x'_{i,j+1}$ with probability $p_f$ (as in the flipping attack). By doing so, the malicious SP (i) distorts the fingerprint with a high probability (by distorting the data points that are not compliant with the inherent correlations in the data) and (ii) adds random noise to the data to further reduce the chance of being detected by the data owner. Following a similar strategy, the malicious SP checks all pairs up to $x'_{i,l-1}$ and $x'_{i,l}$. If the malicious SP does not combine correlation attack with flipping attack, Alice can perform a similar correlation attack on $\mathcal{X}_i'$ and obtain a similar result with the malicious SP. Therefore, the randomness in the flipping attack makes it difficult for Alice to detect the malicious SP.

\subsubsection{Collusion Attack}\label{sec:threat:collusion}

If multiple malicious SPs collude, by comparing their copies, they may detect and distort the fingerprinted data points. The goal of the colluding SPs is to share a single copy of data owner's data without being detected. One well-known collusion attack against fingerprinting schemes, called majority attack in the literature~\cite{li2005fingerprinting}, is when colluding SPs compare all their received data points and choose to share the data value that is observed by the majority of the colluding SPs. However, doing such an attack alone cannot be successful for attackers if there is no randomness in the attack. Otherwise, similar to correlation attack, data owner (Alice) can simulate the majority attack of the colluding SPs (and identify the malicious SPs easily). Therefore, collusion attack should also be combined with flipping attack. Furthermore, the colluding SPs can also use correlations to perform a more successful attack. In Section~\ref{sec:collusion}, we explain how such an attack (i.e., collusion attack combined with correlation and flipping attacks) can be performed against the proposed fingerprinting scheme in detail.

\subsection{Robustness Measures}\label{sec:robustness}

One common requirement of fingerprinting schemes is their robustness against malicious attacks that may destroy or distort the embedded fingerprint. This is important because such attacks may cause the data owner to accuse an innocent SP due to the data leakage, and hence cause false positives in the fingerprint detection algorithm. Therefore, a fingerprinting scheme is considered as robust if it resists malicious attacks and allows the detection of the guilty SP that leaks the data after performing an attack. In this work, we assume that the detection algorithm always returns a guilty SP when Alice observes an unauthorized copy of her data. The proposed detection methods compute a score for each SP having a copy of Alice's data and identify the guilty SP based on these scores. In order to quantify the robustness of the proposed scheme, we use the accuracy ($a$) of the detection algorithm which is defined as the probability of detecting the guilty SP from the leaked data. In~\cite{li2005fingerprinting}, misattribution false hit ($fh^A$) is defined as a robustness measure (defined as the probability of detecting an incorrect fingerprint from the leaked data). 
Therefore, our robustness measure $a$ is equivalent to $(1-fh^A)$. 

If collusion attack occurs, the accuracy of the detection algorithm can be evaluated by the probability of correctly identifying one or all colluding SPs. In~\cite{wu2004collusion}, the possible goals for designing the fingerprinting schemes are defined as ``catch one'', ``catch many'', and ``catch all''. In catch one scenario, the goal is to design the scheme to maximize the probability of catching one of the colluding SPs, while seeking to minimize the probability of falsely accusing an innocent user. In the other scenarios, the probability of falsely accusing the innocent SPs ($fh^A$) increases. Hence, most collusion resistant fingerprinting schemes, such as Boneh-Shaw codes~\cite{boneh1}, aim to catch one of the colluding SPs. Since we utilize Boneh-Shaw fingerprinting codes in our proposed scheme, we also aim to catch one of the colluding SPs in case of collusion attack. Thus, we define $a$ as the probability of detecting one guilty SP from the leaked data.

%% file: sections/algorithm.tex
\section{Probabilistic Fingerprinting Scheme for Correlated Data}
\label{sec:scheme}

In this section, we propose our probabilistic fingerprinting scheme which considers correlations in data (considering the attack in Section~\ref{sec:threat:correlation}). We also present two techniques for the data owner to detect the source of data leakage. In Section~\ref{sec:collusion}, we will improve this scheme to also consider colluding malicious SPs.

\subsection{Proposed Fingerprinting Algorithm}
\label{sec:alg:first}

Assume data owner (Alice) has a sequence of data points $\mathcal{X} = \lbrack x_1, \ldots, x_l \rbrack$ and she wants to share her data with an $SP_i$ as $\mathcal{X}_i' = \lbrack x'_{i,1}, \ldots, x'_{i,l} \rbrack$ after fingerprinting. 
Alice determines a fingerprinting probability $p$ (for each data point), which means approximately $p \cdot l$ data points will be fingerprinted (i.e., their value will be changed) when sharing $l$ data points with $SP_i$. Lower $p$ values increase the utility of shared data (by changing values of fewer data points), however they also decrease the robustness of the fingerprint (i.e., chance of the data owner to detect the source of data leakage). 

Under these settings, a naive algorithm fingerprints each data point with the same probability, without considering correlations in the data. Hence, each data point is shared correctly ($x'_{i,j} = x_{j}$) with probability $1-p$ and incorrectly/fingerprinted ($x'_{i,j} \neq x_{j}$) with probability $p$. For each fingerprinted (incorrectly shared) data point, the shared state is selected among $(m-1)$ states in $\mathcal{D}$ with equal probability. However, if Alice applies this naive probabilistic fingerprinting scheme, a malicious SP can detect some of the fingerprinted data points using the correlations and distort the fingerprint via flipping, as discussed in Section~\ref{sec:threat:correlation}.

In order to prevent such an attack, one needs to consider the correlations in the fingerprinting scheme. In our proposed probabilistic fingerprinting scheme, for each data point $x_j$, considering the correlations in the data, we assign a different probability for sharing each different state of this data point in $\mathcal{D}$. Let $P_{x_j,d_k}$ be the probability of sharing data point $x_j$ as $d_k$ (i.e., $x'_{i,j} = d_k$). The proposed scheme assigns a $P_{x_j,d_k}$ value for all $j \in \{1,2,\ldots,l\}$ and $k \in \{1,2,\ldots,m\}$. By doing so, unlike the naive approach, we consider the correlations in the data, and hence prevent a malicious SP to detect and distort fingerprints.
We propose an iterative algorithm that starts from the first data point $x_1$ and assigns probabilities $P_{x_1,d_1}, \ldots, P_{x_1,d_m}$. Since we consider correlations between consecutive data points, for the first data point $x_1$, similar to the naive approach, Alice shares the correct value with probability $1-p$ and each incorrect value with probability $p/(m-1)$. Based on these probabilities, the algorithm selects a value for $x'_{i,1}$ from the set $\mathcal{D}$. For the subsequent data points, the algorithm computes the fingerprinting probabilities by checking the correlation with the preceding data points. 

Let the shared value of $x_{j-1}$ (i.e., $x'_{i,j-1}$) be $d_{\alpha}$. The algorithm checks the conditional probabilities $P(x_{j} = d_{k} | x_{j-1} = d_{\alpha})$ for all $d_{k} \in \mathcal{D}$. If the algorithm decides to share $x_{j}$ as $d_{k}$, and if $P(x_{j} = d_{k} | x_{j-1} = d_{\alpha})$ is low, a malicious SP can detect that either $x'_{i,j-1}$ or $x'_{i,j}$ is fingerprinted. To eliminate such correlation attack, the proposed algorithm uses a threshold $\tau$ and sets $P_{x_j,d_k} = 0$ if $P(x_{j} = d_{k} | x_{j-1} = d_{\alpha}) < \tau$. This means that the algorithm never selects $d_{k}$ as the value of $x'_{i,j}$ if $d_{k}$ is not consistent with the inherent correlations in the data. Let $d_c$ be the actual value of $x_j$. If $P(x_{j} = d_{c} | x_{j-1} = d_{\alpha}) \geq \tau$, $P_{x_j,d_c}$ is set to $1-p$. All of the remaining probabilities ($P_{x_j,d_k}, k \neq c$) are assigned directly proportional to the value of $P(x_{j} = d_{k} | x_{j-1} = d_{\alpha})$. After assigning all probabilities, the algorithm chooses one of the values from $\mathcal{D}$ based on the assigned probabilities and sets the value of $x'_{i,j}$ accordingly.

Since the proposed algorithm considers correlations between consecutive data points, it may result in changing the value of high number of consecutive data points or sharing the correct value for high number of consecutive data points to prevent malicious SPs from detecting and distorting the fingerprint. As a result of this, total number of fingerprinted points in the original data may significantly deviate from the expected number ($p \cdot l$). For instance, if $p$ is selected as $0.2$, we expect to fingerprint approximately $20\%$ of total data points. However, considering correlations may cause fingerprinting significantly more (or fewer) data points than anticipated. To prevent this, we dynamically decrease (or increase) the fingerprinting probability $p$ depending on number of currently fingerprinted data points as follows.

To keep the average number of fingerprinted data points as $p \cdot l$, the proposed algorithm divides the data points into blocks consisting of $\lceil 1/p \rceil$ data points. We expect (on the average) one fingerprinted data point in each block. Therefore, the algorithm keeps a count of the number of fingerprinted data points at the end of each block. If the ratio of fingerprinted data points is less than $p$, the algorithm sets the fingerprinting probability for the next block as $p\cdot(1 + \theta)$. Here, $\theta$ is a design parameter in the range $\lbrack0,1)$ and we evaluate the selection of $\theta$ in Section~\ref{sec:eval}. If the ratio of fingerprinted data points is greater than $p$, the algorithm sets the fingerprinting probability for the next block as $p\cdot(1 - \theta)$. 

\begin{algorithm}
\small
\SetKwInOut{Input}{input}
\SetKwInOut{Output}{output}
\Input{Original data $\mathcal{X} = \lbrack x_1, x_2, \ldots, x_l \rbrack $, fingerprinting probability $p$, probability adjustment parameter $\theta$, block size $\lceil 1/p \rceil$, correlation threshold $\tau$, pairwise correlations between data points.}
\Output{Fingerprinted copy $\mathcal{X}_i' = \lbrack x'_{i,1}, x'_{i,2}, \ldots, x'_{i,l} \rbrack$}
$ prob \longleftarrow p$\;
      \ForAll{$j \in  \{1,2,\ldots,l\}$}{
        \ForAll{$k \in  \{1,2,\ldots,m\}$}{
            \uIf{$j = 1 ~\&~ x_j = d_k$} { $P_{x_j,d_k} \longleftarrow 1 - prob$;
	     	}
	     	\uElseIf{$j = 1 ~\&~ x_j \neq d_k$} { 
	     	$P_{x_j,d_k} \longleftarrow prob / (m-1)$;
	     	}
	     	\uElseIf{$P(x_{j} = d_{k} | x_{j-1} = x'_{i,j-1}) < \tau$}{
	     	$P_{x_j,d_k} \longleftarrow 0$;
	     	}
	     	\uElseIf{$x_{j} = d_{k}$}{
	     	$P_{x_j,d_k} \longleftarrow 1 - prob$;
	     	}
        }
        distribute the remaining probability (1 - (sum of assigned probabilities)) by assigning $P_{x_j,d_k}$ directly proportional to the value of $P(x_{j} = d_{k} | x_{j-1} = x'_{i,j-1})$ if $P_{x_j,d_k}$ is not assigned in the previous step;
        
        $x'_{i,j} \longleftarrow$ random value from $\mathcal{D}$ using probability distribution $P_{x_j,d_1},...,P_{x_j,d_m}$;
        
        \If{$j$ is multiple of $\lceil 1/p \rceil$} { $c\longleftarrow $ total number of fingerprinted data points\;
       \uIf{$c > p \cdot j$} { $prob \longleftarrow p \cdot(1 - \theta)$;
	     		}\uElseIf{$c < p \cdot j$} { $prob \longleftarrow p \cdot(1 + \theta)$;
	     		}\Else{$prob \longleftarrow p$;}
	     		}
     }
		\caption{Probabilistic fingerprinting scheme without considering the collusion attack.}
\label{alg:alg1}
\end{algorithm}

The steps of the proposed scheme are also shown in Algorithm~\ref{alg:alg1}. For each SP, Alice executes the same algorithm with a different seed value and stores the fingerprint pattern of each SP to use it in the detection in case her data is shared without her consent. If the data size and number of SPs are large, Alice can also just store the seed value for each SP.

\begin{figure}
\centering
\includegraphics[width=13cm,keepaspectratio]{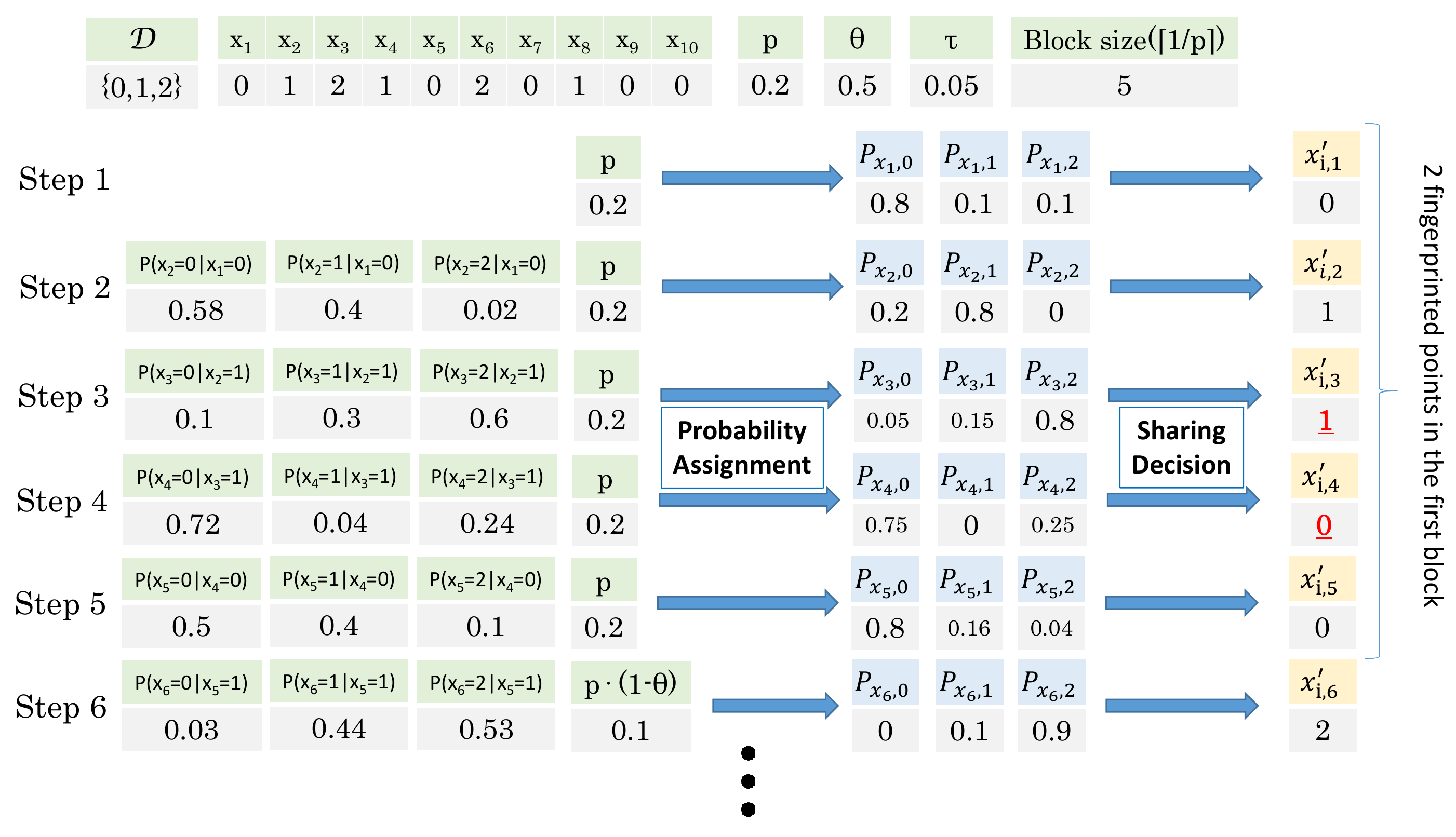}
  \vspace{-5mm}
\caption{A toy example showing the execution of Algorithm~\ref{alg:alg1}. Input parameters of the algorithm are shown at the top as the original data $\mathcal{X}$, the fingerprinting probability $p$, probability adjustment parameter $\theta$, correlation threshold $\tau$, and the block size $\lceil 1/p \rceil$. Data values with correlation less than $\tau$ are assigned with $0$ probability. }
  \vspace{-3mm} 
\label{fig:algExample}
\end{figure} 

Figure~\ref{fig:algExample} shows a toy example to illustrate the execution of the proposed algorithm (Algorithm~\ref{alg:alg1}). Each step shows one execution of the outer loop of Algorithm~\ref{alg:alg1} for deciding the value of one data point. In the first step, probabilities are assigned just by using $p$. In the next steps, correlation values are also used to determine the probabilities that are used to choose the shared values ($x'_{i,j}$). When the correlation is less than $\tau = 0.05$, the algorithm assigns $0$ for the probability of selecting the corresponding value. At the end of each block (a block includes $5$ data points in the example), the algorithm checks the total number of fingerprinted values. Since the expected number of fingerprinted points in a block is $1$ and the algorithm fingerprinted $2$ data points, the fingerprinting probability is adjusted as $p \cdot (1 - \theta) = 0.1$ for the second block.

\subsection{Detecting the Source of Data Leakage}
\label{sec:single:detection}

Let the leaked copy of Alice's data be $\mathcal{Y} = \lbrack y_1, \ldots, y_l \rbrack$. The goal of Alice is to detect the SP that leaks her data. Here, we propose two methods that can be used by Alice to detect the source of the data leakage. Both methods compute a score (probability or similarity) for each SP (with whom Alice previously shared her data) to be guilty for leaking $\mathcal{Y}$. Alice chooses the SP that has the highest score as the guilty SP. Note that, the size of the leaked data may not be equal to the size of the Alice's data (i.e., $l$) if subset attack is performed. In that case, we represent excluded data points with $d_0$ (that is not in set $\mathcal{D}$).   

\subsubsection{Probabilistic Detection} \label{sec:probDetect}

In~\cite{papadimitriou2010data}, considering the data leakage problem, the authors compute the probability of being guilty for each agent (SP) with some independence assumptions (to efficiently compute the probabilities). We adapt their detection method to our problem setting as follows: For each leaked data point $y_j$, Alice identifies each SP $i$ that received the corresponding data point as $y_j$ (i.e., $x'_{i,j} = y_j$). Thus, for each data point $y_j$, Alice constructs a set $\mathcal{V}_j$ consisting of such SPs (that received the leaked state of the data). Alice assumes that each SP in $\mathcal{V}_j$ can be guilty for leaking $y_j$ with a probability $1/|\mathcal{V}_j|$. For instance, if the value of $y_j$ is shared with $4$ SPs, each of these SPs is considered as guilty for sharing $y_j$ with probability $0.25$. Eventually, an $SP_i$ is considered as guilty with probability $1 - \prod_{ y_j \in \mathcal{Y} , y_j = x'_{i,j}} (1 - \frac{1}{|\mathcal{V}_j|})$. At the end, Alice identifies the guilty SP as the one with highest probability of being guilty.

\subsubsection{Similarity-Based Detection} \label{sec:similarity}

Another efficient way of detecting the guilty SP is computing the similarity of fingerprinted data shared with each SP with the leaked data. For an $SP_i$, Alice compares the leaked data $\mathcal{Y}$ with the copy $\mathcal{X}_i'$ and counts the matching data points in the fingerprint pattern. In other words, Alice checks the size of the following set: $\mathcal{M}_i = \{x_j~|~x_j \in \mathcal{X},~x_j \neq x'_{i,j},~x'_{i,j} = y_j \}$. Alice also counts the fingerprinted data points $\mathcal{F}_i = \{x_j~|~x_j \in \mathcal{X},~x_j \neq x'_{i,j} \}$ of $SP_i$. Eventually, the SP with the maximum $sim_i = |\mathcal{M}_i|/|\mathcal{F}_i|$ value is identified as guilty.

Similarity-based detection is easy to implement and it efficiently detects the guilty SP if there is no collusion between the SPs or the number of colluding SPs is small. However, when the number of colluding SPs increases, the similarity scores of SPs become very close to each other and the detection performance of similarity-based detection decreases. Probabilistic detection differs from similarity-based detection by considering the number of SPs having the leaked data point (i.e., $|\mathcal{V}_j|$). In similarity-based detection, the effect of each data point to the similarity score is same. However, in probabilistic detection, the probability of being guilty is computed by multiplying $(1 - \frac{1}{|\mathcal{V}_j|})$ values, and hence the effect of each data point to the probability of being guilty is different. In Section~\ref{sec:eval}, we compare the performance of these two detection techniques in terms of their accuracy to identify the guilty SP.

%% file: sections/collusions.tex
\section{Considering Colluding Service Providers}
\label{sec:collusion}

As discussed in Section~\ref{sec:threat:collusion}, the colluding SPs may distort the fingerprint via the majority attack. However, lack of randomness in this attack may result in detection of colluding SPs by Alice. Here, we first show a strong attack against the proposed fingerprinting scheme (also against the existing fingerprinting schemes in general) by integrating colluding SPs, correlations in the data (in Section~\ref{sec:threat:correlation}), and the flipping attack (in Section~\ref{sec:threat:flipping}). Then, we propose utilizing Boneh-Shaw codes~\cite{boneh1} to improve robustness against such a strong attack that also involves colluding SPs.

\vspace{-1mm}
\subsection{Probabilistic Majority Attack}
\label{sec:collusionattack}

As discussed, the goal of the colluding SPs is to share (leak) a copy of the data without being detected by the data owner (Alice). In a standard collusion attack, the colluding SPs compare their received values for each data point and select the most observed value to share. We propose an advanced collusion attack called ``probabilistic majority attack'' (to distinguish it from the standard majority attack), in which the colluding SPs decide the value of each leaked data point by considering (i) all observed values for that data point, (ii) correlation of that data point with the others, and (iii) the probability of adding a fingerprint to a data point ($p$). If the colluding SPs do not know the fingerprinting probability $p$, we assume they use an estimated probability $p_e$ in their attack ($p_e=p$ if $p$ is publicly known). 

Let the set of colluding SPs be $\mathcal{C}$ and $|\mathcal{C}|=n$. The goal of the colluding SPs is to create a copy $\mathcal{Y} = \lbrack y_1, y_2, \ldots, y_l \rbrack$ to share and avoid being detected by Alice. In this attack, the colluding SPs first decide the value of each data point $y_j$ by computing a probability $P_{y_j,d_k}$ for each possible state $d_k \in \mathcal{D}$ of $y_j$. To compute $P_{y_j,d_k}$, colluding SPs first check their received values for data point $j$. Let $c_{j,d_k}$ be the number of observations of $d_k$ for data point $j$ by $n$ colluding SPs (in $\mathcal{C}$). In the standard majority attack, the colluding SPs choose $d_k$ with the maximum $c_{j,d_k}$ value as the value of $y_j$ (assuming it is the original value of the data point) to avoid detection by Alice. However, it is possible (with lower probability) that other values with lower $c_{j,d_k}$ may also be the original value of $y_j$. For example, assume there are three colluding SPs and let $\mathcal{D} = \{d_1, d_2, d_3\}$. For a data point $j$, observing two $d_1$s ($c_{j,d_1} = 2$) and one $d_2$ ($c_{j,d_2} = 1$) is not enough to conclude that $d_1$ is the original value ($x_j$) of data point $j$. It is also possible (with probability $(p/2)^3$) that the original value is $d_3$ and all three colluding SPs received fingerprinted (noisy) values from Alice.

Thus, in the probabilistic majority attack, the attackers compute such probabilities for each $d_k \in \mathcal{D}$ using their estimated fingerprinting probability $p_e$ and also using publicly known pairwise correlations between the data points. As discussed in Section~\ref{sec:threat:correlation}, correlations can be used to detect and distort fingerprinted data points by the attackers. Therefore, the attackers tend to select each leaked value that have high correlation with the previous shared (leaked) values. In order to integrate the correlations with the collusion attack, the conditional probabilities (due to correlations) are used as weights to determine the sharing probabilities of colluding SPs ($P_{y_j,d_k}$). The colluding SPs first compute the weighted probability values (referred as $t_{j,d_k}$) and then compute $P_{y_j,d_k}$ by normalizing the weighted probability values. The weighted probability for $d_k$ value of a data point $j$ is computed as follows: 
\begin{displaymath}
t_{j,d_k} = (1-p_e)^{c_{j,d_k}} \cdot (\frac{p_e}{m-1})^{n-{c_{j,d_k}}} \cdot P(x_j = d_k | x_{j - 1} = y_{j - 1}).
\end{displaymath}
Here, $(1-p_e)^{c_{j,d_k}} \cdot (\frac{p_e}{m-1})^{n-{c_{j,d_k}}} $ is the probability of $d_k$ to be the original value of data point $j$ by assuming each data point is fingerprinted with probability $p_e$. The conditional probability $P(x_j = d_k | x_{j - 1} = y_{j - 1})$ is used as a weight. Then, $P_{y_j,d_k}$ is computed by normalizing $t_{j,d_k}$ values as 
\begin{displaymath}
P_{y_j,d_k} = t_{j,d_k}/(\sum_{k=1}^{n} t_{j,d_k}).
\end{displaymath}
The colluding SPs decide on the value of each shared point $y_j$ proportional to these probabilities. Therefore, the malicious SPs do not necessarily select the value observed by the majority. By doing so, we allow the malicious SPs to further distort the fingerprint compared to the standard majority attack.

Furthermore, existing techniques to prevent collusion attacks (e.g., Boneh-Shaw codes) assign common fingerprints to multiple SPs, which allows data owner to detect colluding SPs with a high chance. However, to avoid such detection, the attackers may flip some random data points before they leak the data. Thus, in the probabilistic majority attack, we also let colluding SPs flip each $y_j$ with probability $p_f$ (as discussed in Section~\ref{sec:threat:flipping}).

To illustrate the probabilistic majority attack with a toy example, let $n=4$ and $\mathcal{D} = \{0,1,2\}$. Assume $3$ of the colluding malicious SPs have received value $0$ for the first data point ($x_1$) and the other malicious SP has received value $1$. Let the estimated fingerprinting probability $p_e=0.1$. Colluding SPs compute $t_{1,0} = (0.9)^{3} \cdot (0.1)^{1} \cdot 1$, $t_{1,1} = (0.9)^{1} \cdot (0.1)^{3} \cdot 1$, and $t_{1,2} = (0.9)^{0} \cdot (0.1)^{4} \cdot 1$. Since $x_1$ is the first data point and we consider pairwise correlations between consecutive data points, here, conditional probabilities are all considered as $1$. Then, colluding SPs choose a value (to share) from $\mathcal{D}$ with the following probabilities: $P_{y_1,0} = 0.0729 / 0.0739 = 0.987$, $P_{y_1,1} = 0.0009 / 0.0739 = 0.012$, and $P_{y_1,2} = 0.0001 / 0.0739 = 0.001$. Finally, the chosen value is flipped with probability $p_f$. 

\subsection{Integrating Boneh-Shaw Codes}
\label{sec:integrate-boneh}

In order to provide robustness against collusion attacks, Boneh and Shaw proposed fingerprinting codes for detecting one of the colluding SPs~\cite{boneh1}. The effectiveness of their codes depend on the ``marking assumption'', which states that colluding SPs cannot detect the fingerprint if all of them have the same fingerprint. Hence, when the colluding SPs have the same value for the same data point $j$, it is assumed that they choose this value as $y_j$ as the leaked data point. However, Boneh-Shaw codes do not consider correlation and flipping attacks, and hence their detection method is not successful when colluding SPs also utilize the correlations in the data and use the flipping attack. Although Boneh-Shaw codes are defined for binary data, we apply these codes to non-binary data. We briefly describe the Boneh-Shaw codes in Section~\ref{sec:boneh-shaw} and show how to utilize these codes in the proposed scheme in Section~\ref{sec:use-boneh-shaw}. 
\subsubsection{Boneh-Shaw Codes} \label{sec:boneh-shaw}

Boneh-Shaw codes~\cite{boneh1} are designed to detect one of the colluding SPs with probability at least $(1 - \omega)$, where at most $c$ SPs can collude, under the marking assumption (in the original paper~\cite{boneh1}, the error probability is represented as $\epsilon$, here we use $\omega$ to prevent confusion with privacy parameter $\epsilon$ in differential privacy). Note that $c$ is the maximum number of colluding SPs, for which the detection technique can provide robustness (i.e., detect one of the colluding SPs) with high probability ($\omega$-error) and $n$ represents the actual number of colluding SPs in the collusion attack. Boneh-Shaw codes are defined as binary codes, in which each codeword consists of zeros and ones. In order to fingerprint data, some data points are selected from the original data and XOR'ed with the permuted codeword in order to prevent colluding SPs from detecting and distorting the fingerprint (the number of data points must be equal to the length of the codeword). The same permutation must be used for each SP and the permutation must be hidden from SPs. The data points that are XOR'ed with ones in the codeword becomes fingerprinted. Therefore, the ones in the binary code represent fingerprints. 

In Boneh-Shaw ($c$,$r$)-codes, the first codeword consists of $(c-1) \cdot r$ ones, $i$th codeword consists of $(i-1) \cdot r$ zeros and $(c-i)\cdot r$ ones, and the last ($c$th) codeword consists of $(c-1) \cdot r$ zeros. For instance, the four codewords of $(4,3)$-codes are $[111111111]$, $[000111111]$, $[000000111]$, and $[000000000]$. If the recipients of first and third codewords collude, based on the marking assumption, they create $[RRRRRR111]$ (as the leaked data), where $R$ is a randomly selected bit. To detect the source of the leakage (i.e., the guilty SP), the data owner checks each $r$-bit block, observes that the third block consists of all ones and the second block contains a zero with high probability, and concludes that the recipient of the third codeword is involved in the collusion. However, it is also possible that the colluding SPs create $[010111111]$, which leads the data owner to accuse the owner of the second codeword. Thus, increasing $r$ decreases the error in detection ($\omega$), but it also increases the length of the fingerprint (and hence decreases the utility of the shared data). 

\subsubsection{Using Boneh-Shaw Codes in the Proposed Scheme}\label{sec:use-boneh-shaw}

The marking assumption of~\cite{boneh1} does not consider the flipping attack and correlation attack which are included in the probabilistic majority attack. If the colluding SPs flip some of the bits randomly or based on correlations, the data owner may accuse an SP who is not involved in the collusion. Therefore, it is not possible to directly use Boneh-Shaw codes and their detection algorithm in our scenario. Instead, we utilize Boneh-Shaw codes to assign shared (i.e., overlapping) fingerprints between different SPs. 

In Algorithm \ref{alg:alg1}, Alice creates fingerprinted copy of each SP independently. Hence, the fingerprints of two SPs may be the same for some random data points. Here, we explicitly assign overlapping fingerprints using Boneh-Shaw codes to improve robustness against collusion attacks. In order to adopt binary codes into our scheme, we consider ones as fingerprinted data points and zeros as the original data points. Therefore, when a data point is fingerprinted using Boneh-Shaw codes, the same value is also used as a fingerprint in the copies for other SPs if their codewords also include one for the same data point. 

We integrate the Boneh-Shaw codes into our scheme as follows: Alice creates the first fingerprinted copy of her data $\mathcal{X}_1'$ (to share with $SP_1$) as described in Algorithm~\ref{alg:alg1}. Approximately $p \cdot l$ data points are fingerprinted in $\mathcal{X}_1'$. Let $f$ be the number of fingerprinted data points for $SP_1$. We want to use some portion of these $f$ data points as Boneh-Shaw codes. Then, Alice decides the value of $c$ and $r$ such that $(c-1) \cdot r \leq f$ to apply Boneh-Shaw codes for the next sharings. As mentioned, $c$ and $r$ are design parameters of Boneh- Shaw codes determining the length of codes and error in detection ($\omega$). We represent $(c-1) \cdot r$ as $f_{1}$, which is the length of Boneh-Shaw codeword. Here, $c$ is the number of Boneh-Shaw codewords that Alice can create and $r$ is the block size, as explained in Section~\ref{sec:boneh-shaw}. If Alice wants to share her data with more than $c$ SPs, she assigns the same Boneh-Shaw codewords in a similar order. $SP_{c+1}$ receives the same codeword with $SP_1$, $SP_{c+2}$ receives the same codeword with $SP_2$, and so on. In this way, although same Boneh-Shaw codewords are assigned to some SPs, since other parts of their fingerprints will be different, the proposed detection algorithm can still identify the guilty SP using the entire fingerprint. If $f_1 = f$, Alice uses all fingerprint of $SP_1$ for Boneh-Shaw codes in later sharings of her data (with other SPs). This provides better robustness against collusion attacks, since the number of overlapping fingerprints is high. However, in this case, the robustness for the attacks performed by single SP (e.g., flipping or correlation attack) will become weak because the number of unique fingerprints is low. Therefore, we set $f_1$ approximately equal to $f/2$ to detect the guilty SP even the attack is performed by single SP or multiple SPs.  

Alice randomly selects $f_1$ of $f$ fingerprinted points in $\mathcal{X}_1'$. These $f_1$ fingerprinted data points are considered as the first codeword in Boneh-Shaw codes. For her next sharing with the second SP ($SP_2$), Alice randomly selects $f_2 = f_1 - r$ of $f_1$ points to assign the same fingerprints to $SP_2$ (i.e., $x'_{2,j} = x'_{1,j}$ for these $f_2$ data points). Furthermore, Alice, assigns the original value for the remaining $r$ points (i.e., $x'_{2,j} = x_{j}$ for these $r$ data points). In other words, the Boneh-Shaw codeword of $SP_2$ consists of $r$ zeros and $f_1 - r$ ones. 

In order to assign approximately $p \cdot l$ fingerprinted points to $SP_2$, the fingerprinting probability of $SP_2$ is selected as $\frac{p\cdot l - f_2}{l - f_1}$ since $f_2$ fingerprints are already assigned before running the probabilistic algorithm. Alice runs the proposed algorithm to sequentially add the remaining fingerprints. Also, since $f_2$ fingerprints and $r$ original values are already assigned (as the Boneh-Shaw codeword of $SP_2$) before the algorithm, the algorithm will skip these points while adding fingerprints. Furthermore, when the algorithm is determining the probabilities for each possible value of a data point (i.e., inner loop of the algorithm), it also considers the correlations of the data points with the already assigned Boneh-Shaw codeword. The updated algorithm is shown in Algorithm~\ref{alg:alg2}, which includes these new conditions. For each $SP_i$, Alice repeats the same process by first adding $f_i$ fingerprints and $f_1 - f_i$ original values (i.e., Boneh-Shaw codeword). Then, Alice runs the probabilistic algorithm to determine the values of remaining points in  $\mathcal{X}_i'$ as in Algorithm~\ref{alg:alg2}. 

\begin{algorithm}[t]
\small
\SetKwInOut{Input}{input}
\SetKwInOut{Output}{output}
\Input{Original data $\mathcal{X} = \lbrack x_1, x_2, \ldots, x_l \rbrack $, \color{blue}$f_1$ already assigned points in $\mathcal{X}_i'$ ($f_i$ of them are fingerprinted), fingerprinting probability $\frac{p\cdot l - f_i}{l - f_1}$\color{black}, probability adjustment parameter $\theta$, block size $\lceil 1/p \rceil$, correlation threshold $\tau$, pairwise correlations between data points.}
\Output{Fingerprinted copy $\mathcal{X}_i' = \lbrack x'_{i,1}, x'_{i,2}, \ldots, x'_{i,l} \rbrack$}
$ prob \longleftarrow \color{blue}\frac{p\cdot l - f_i}{l - f_1}$\;\color{black}
      \ForAll{$j \in  \{1,2,\ldots,l\}$}{
      \color{blue}\If{$x'_{i,j}$ is not assigned}{
      \color{black}
        \ForAll{$k \in  \{1,2,\ldots,m\}$}{
            \uIf{$j = 1 ~\&~ x_j = d_k$} { $P_{x_j,d_k} \longleftarrow 1 - prob$;
	     	}
	     	\uElseIf{$j = 1 ~\&~ x_j \neq d_k$} { 
	     	$P_{x_j,d_k} \longleftarrow prob / (m-1)$;
	     	}
	     	\uElseIf{$P(x_{j} = d_{k} | x_{j-1} = x'_{i,j-1}) < \tau$}{
	     	$P_{x_j,d_k} \longleftarrow 0$;
	     	}
	     	\color{blue}\uElseIf{$x'_{i,j+1}$ is assigned \& $P(x_{j+1} = x'_{i,j+1} | x_{j} = d_{k}) < \tau$}{
	     	$P_{x_j,d_k} \longleftarrow 0$;
	     	}\color{black}
	     	\uElseIf{$x_{j} = d_{k}$}{
	     	$P_{x_j,d_k} \longleftarrow 1 - prob$;
	     	}
        }
        distribute the remaining probability (1 - (sum of assigned probabilities)) by assigning $P_{x_j,d_k}$ directly proportional to the value of $P(x_{j} = d_{k} | x_{j-1} = x'_{i,j-1})$ if $P_{x_j,d_k}$ is not assigned in the previous step;
         
         $x'_{i,j} \longleftarrow$ random value from $\mathcal{D}$ using probability distribution $P_{x_j,d_1},...,P_{x_j,d_m}$;
        }
        \color{black}
          \If{$j$ is multiple of $\lceil 1/p \rceil$} { $c\longleftarrow $ total number of fingerprinted data points\;
       \uIf{$c > p \cdot j$} { $prob \longleftarrow p \cdot(1 - \theta)$;
	     		}\uElseIf{$c < p \cdot j$} { $prob \longleftarrow p \cdot(1 + \theta)$;
	     		}\Else{$prob \longleftarrow p$;}
	     		}

     }
		\caption{Probabilistic fingerprinting scheme after assigning the Boneh-Shaw codeword to provide robustness against the collusion attack. Blue parts represent the difference with Algorithm~\ref{alg:alg1}.}
\label{alg:alg2}
\end{algorithm}

\begin{figure}[ht]
\centering
\includegraphics[width=13cm,keepaspectratio]{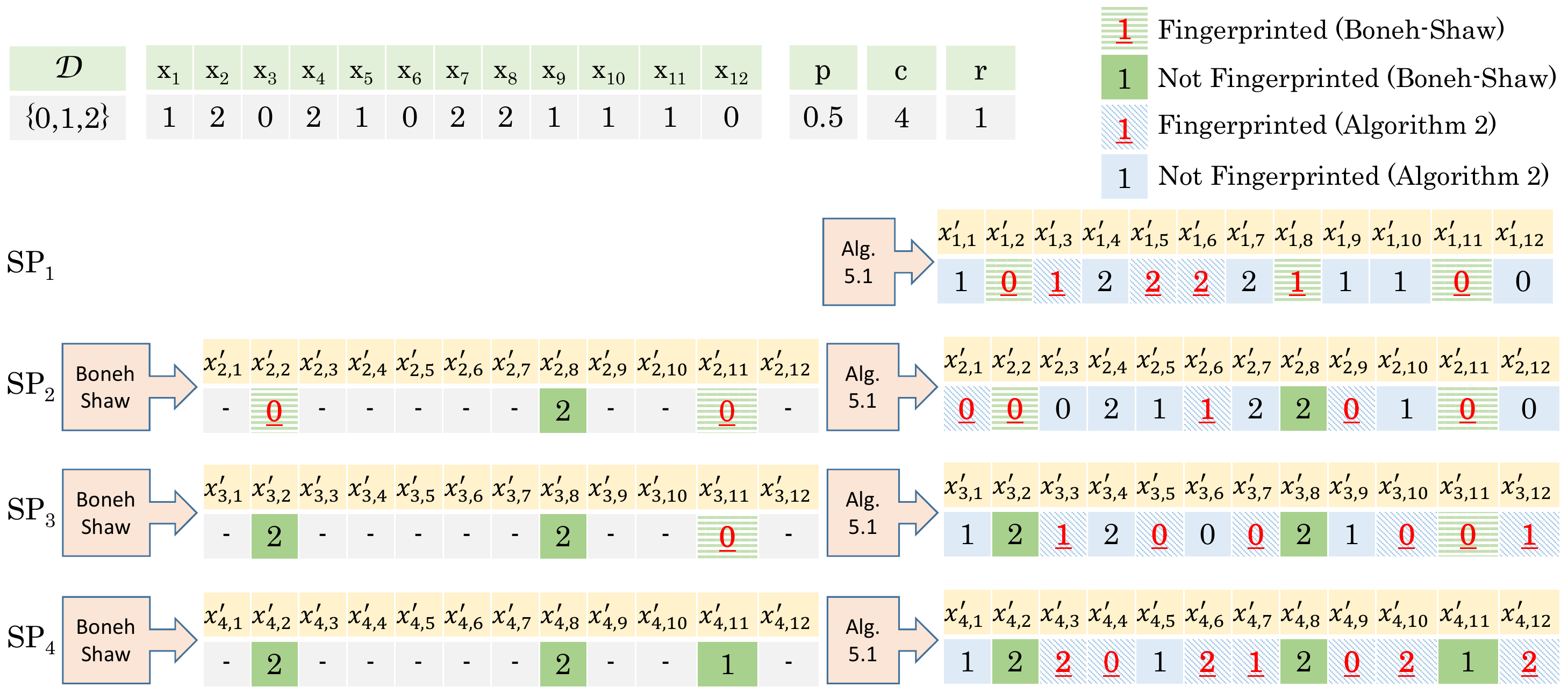}
  \vspace{-2mm}
\caption{An example execution of the proposed algorithm by integrating Boneh-Shaw codes.}
\label{fig:colExample}
\end{figure} 

Here, we describe the algorithm explained in Section~\ref{sec:integrate-boneh} on a toy example, which is also illustrated in Figure~\ref{fig:colExample}. Let the original data of Alice be $\mathcal{X} = \lbrack 1,2,0,2,1,0,2,2,1,1,1,0 \rbrack$, where $l = 12$ and $\mathcal{D} = \{0,1,2\}$. Let $p$ be $0.5$ and Alice shares $\mathcal{X}_1' = \lbrack 1, \underline{0}, \underline{1}, 2, \underline{2}, \underline{2}, 2, \underline{1}, 1,$ $1, \underline{0}, 0 \rbrack$ with $SP_1$ after running Algorithm~\ref{alg:alg2}, where underlined points represent fingerprinted data points.
We use such a high $p$ value for the clarity of this toy example, typically the value of $p$ is significantly smaller to have higher data utility for Alice. 
Let $c=4$ and $r=1$. Alice selects $f_1 = 3$ random fingerprints from $SP_1$ as the Boneh-Shaw codeword. Let the indices of these selected fingerprints be $2$, $8$, and $11$. For the second SP ($SP_2$), Alice keeps two of these fingerprints and sets the other one as the original value of the data point. Hence, Alice inserts Boneh-Shaw codeword as $\mathcal{X}_2' = \lbrack -,\underline{0},-,-,-,-,-,2,-,-,\underline{0},- \rbrack$ before running Algorithm~\ref{alg:alg2}. Since Alice has already two fingerprinted points, she sets fingerprinting probability for $SP_2$ as $4/9$ in Algorithm~\ref{alg:alg2}. For the third and fourth SPs ($SP_3$ and $SP_4$), Alice inserts the Boneh-Shaw codewords before running Algorithm~\ref{alg:alg2} as $\mathcal{X}_3' = \lbrack -,2,-,-,-,-,-,2,-,-,\underline{0},-\rbrack$ and $\mathcal{X}_4' = \lbrack -,2,-,-,-,-,-,2,-,-,1,-\rbrack$, respectively. Alice assigns the values of missing points (represented with dash) sequentially using the Algorithm~\ref{alg:alg2} as shown in Figure~\ref{fig:colExample}.

\subsection{Detection Algorithm}
\label{sec:collusion:detection}

Here, we propose a detection algorithm for the proposed collusion-resilient fingerprinting scheme. In practice, when Alice realizes that a copy $\mathcal{Y}$ of her data is leaked without her consent, she cannot know whether $\mathcal{Y}$ is leaked by single SP (by performing flipping or correlation attack) or multiple SPs (by performing collusion attack). Therefore, Alice cannot use the detection techniques in Section~\ref{sec:single:detection} or the detection technique of Boneh-Shaw codes directly. We propose a detection algorithm that utilizes both techniques. As we show in the experimental evaluations, similarity-based detection in Section~\ref{sec:single:detection} performs slightly better than probabilistic detection. Hence, here, we describe the detection algorithm using similarity-based detection. It can also be implemented using probabilistic detection technique similarly. 

First, Alice needs to determine whether her data is leaked by single SP or multiple SPs. In order to do so, she initially computes a similarity score ($sim_i$) for each $SP_i$ (that received her data) as explained in Section~\ref{sec:similarity}. Each score is a number in the range $\lbrack0,1 \rbrack$. If there is a collusion of two SPs and the colluding SPs observe different values for a data point, they select either of these values with equal probabilities. Thus, we expect that they damage approximately half of the fingerprinted data points (this will be more if the collusion includes more than 2 SPs). Hence, we assume that there is an attack by single SP if the similarity score of an SP is greater than $0.5$. In such a case, Alice identifies the SP with the highest similarity score as guilty. Otherwise, there is a collusion attack with high probability, and hence Alice identifies the suspects according to their similarity scores and returns one of them utilizing the detection technique of Boneh-Shaw codes. 

Let the index of SP with maximum similarity score be $max$. Alice generates a suspect list by including $\lfloor 1/sim_{max} \rfloor$ SPs having highest similarity scores. Hence, if $sim_{max}$ is greater than $0.5$, there will be just one SP in the suspect list and the algorithm will return $SP_{max}$ as guilty. In such case, Alice concludes that there is no collusion attack with high probability. If $sim_{max}$ is less than or equal to 0.5, there will be more than one suspects, which means that a collusion attack is performed with high probability. In this case, the algorithm returns one of the suspects using the detection method of Boneh-Shaw codes~\cite{boneh1} as follows: In Boneh-Shaw codes, it is expected that the colluding SPs create a copy consisting of several random values followed by all ones and the starting point of ones (a block with all ones) gives us one of the colluding SPs (ones in the Boneh-Shaw codes represent the fingerprinted data points and zeros represent the points that are not fingerprinted). Let $B_R$ represents a block (consists of $r$ data points) having at least one zero value and $B_1$ represents a block having all ones. Assuming the leaked copy created by colluding SPs is $\lbrack B_R,...,B_R,B_1,...,B_1\rbrack$, $SP_i$ is identified as guilty if the first observed $B_1$ block is the $i$th block. However, as a result of probabilistic majority attack and flipping attack described in Section~\ref{sec:collusionattack}, some ones may turn into zeros and some zeros may turn into ones. Therefore, the detection algorithm may fail against such an attack. 
To avoid this, we define $\hat{B}_1$ as a block having majority of points as one and $\hat{B}_R$ as a block having majority of points as zero. Then, the algorithm checks all suspects in the suspect list starting from $SP_{max}$ (having the highest similarity score). If $(max)$th block is $\hat{B}_1$ and $(max-1)$th block is $\hat{B}_R$, the algorithm returns $SP_{max}$ as guilty. Otherwise, the algorithm continues with the other SPs in the suspect list in the order of decreasing similarity scores. For each $SP_i$ in the suspect list, the algorithm returns it as guilty if the $(i)$th block is $\hat{B}_1$ and the $(i-1)$th block is $\hat{B}_R$. When such an SP is found, the algorithm stops and returns the SP as guilty. If there is no such an SP, the algorithm returns $SP_{max}$ as guilty.

\begin{figure}
\centering
\includegraphics[width=10cm,keepaspectratio]{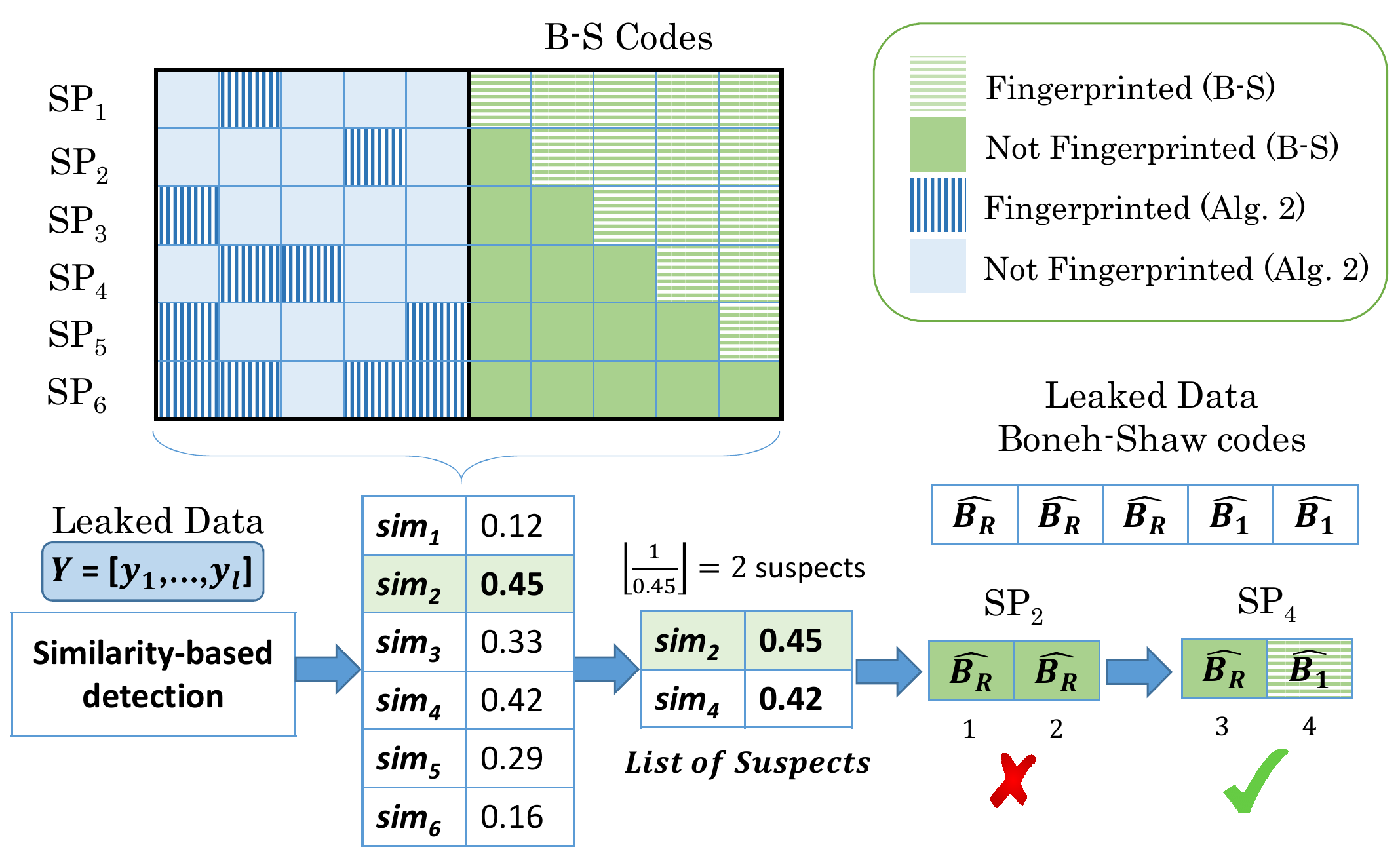}
  \vspace{-2mm}
\caption{An example execution of the detection algorithm.}
  \vspace{-2mm} 
\label{fig:colDetection}
\end{figure}  

The steps of the proposed detection algorithm are also shown with an example in Figure~\ref{fig:colDetection}. After checking the similarity of leaked data with the fingerprinted data points of each SP, the algorithm adds $\lfloor 1/sim_{max} \rfloor = 2$ SPs into the suspect list. When the algorithm checks the $1$st and the $2$nd blocks of leaked data for $SP_2$, it does not return $SP_2$ as guilty since both blocks are $\hat{B}_R$. Then, it checks the $3$rd and the $4$th blocks of leaked data for $SP_4$ and returns $SP_4$ as guilty since the $4$th block is $\hat{B}_1$ and the $3$rd block is $\hat{B}_R$.

%% file: sections/privacy.tex
\section{Using Fingerprinting for Privacy-Preserving Data Sharing}\label{sec:privacy}

Here, we explore how the proposed fingerprinting technique can also provide privacy-preserving data sharing guarantees for the data owner. We develop a mechanism in which the added fingerprint (to provide liability) can also used to provide privacy. When sharing data with untrusted parties, local differential privacy (LDP) is a commonly used concept for providing (statistical) privacy guarantees~\cite{erlingsson2014rappor, wang2017locally}. LDP-based mechanisms typically add controlled noise to data points to guarantee that an untrusted data collector cannot determine the original value of a data point from the reported (perturbed) value. Thus, one trivial idea is to use the noise pattern added by the LDP-based mechanism as the fingerprint. However, one major challenge about this idea is the conflicting objectives of fingerprinting and privacy mechanism.

To show these conflicting objectives via an example, assume that a data owner shares her data with multiple SPs using an LDP-based mechanism by adding noise to different data points based on the LDP parameter ($\epsilon$) at each sharing. In that case, if two or more SPs collude, they can recover (infer the actual values of) most of the noisy data points by just aggregating their received data and selecting the data values that are observed by the majority of them (i.e., similar to performing a standard majority attack). Thus, such a sharing strategy is not preferred for the privacy objective. To overcome this problem, the data owner can just choose to add noise to her data once and share the same noisy data with all the SPs. On the other hand, a fingerprinting mechanism requires the data owner to share different fingerprint patterns with the SPs to uniquely identify the source of an unauthorized sharing. This implies that the robustness of the fingerprinting scheme and the privacy of the shared data are inversely proportional. 

In order to consider both privacy and liability at the same time, we propose a hybrid approach as follows: When Alice (data owner) shares her data with the first SP using the proposed algorithm, she selects a parameter $\lambda$ in the range ($\lbrack0,1\rbrack$) to determine her privacy level in the fingerprinting scheme. She randomly selects $\lambda \cdot f$ of $f$ fingerprinted points as the overlapping fingerprints. These fingerprinted points are shared the same with all the SPs. Thus, for each new data sharing, Alice first inserts $\lambda \cdot f$ overlapping fingerprints and then applies the proposed fingerprinting algorithm for the remaining fingerprints. While higher $\lambda$ values provide higher privacy, lower $\lambda$ values provide better fingerprint robustness. In Section~\ref{sec:ev:privacy}, we show this trade-off between fingerprint robustness and privacy for different values of $\lambda$.

%% file: sections/evaluation.tex
\section{Evaluation}\label{sec:eval}

To evaluate its robustness and utility, we implemented the proposed fingerprinting algorithm (in Section~\ref{sec:integrate-boneh}) and the detection algorithm (in Section~\ref{sec:collusion:detection}). We here present our experimental evaluations. 

\subsection{Data Model and Settings} \label{sec:ev:data}

We evaluated the proposed scheme on a genomic data sharing scenario. Due to the fast decrease in the cost of sequencing, nowadays, individuals can obtain their genome sequences easily. This trend has also led individuals to share their genomic data with medical institutions and direct-to-consumer service providers for various genetic tests or research purposes. Since genomic data contains sensitive personal information, such as the risk of developing particular diseases, sharing genomic data without the authorization of the data owner causes privacy violations. Therefore, fingerprinting genomic data can be a solution or disincentive to prevent its unauthorized sharing. Furthermore, genomic data contains inherent pairwise correlations between point mutations (single nucleotide polymorphisms - SNPs), which makes genomic data an ideal usecase to evaluate the proposed scheme. SNP is the variation of a single nucleotide (from the set $\{A,T,C,G\}$) in the population. For each SNP position, only two different nucleotides can be observed: (i) major allele, which is observed in the majority of the population and (ii) minor allele, which is observed rarely. Moreover, each SNP consists of two nucleotides, one is inherited from the mother and the other from the father. Thus, each SNP is represented by the number of its minor alleles, and $\mathcal{D} = \{0,1,2\}$ for genomic data. 

We used a dataset that consists of 7690 SNPs belonging to 99 people from Central European ethnicity~\cite{site:impute}. Using this dataset, we computed the pairwise correlations between SNPs to build our correlation model. Unless stated otherwise, we set the data size $l =$ 1,000, fingerprinting probability $p = 0.1$, and the correlation threshold of the algorithm $\tau = 0.05$. The threshold $\tau$ is used in the algorithm to prevent adding fingerprints causing low correlation. Note that the attacker has its own correlation threshold $\tau_c$ in its correlation attack. We evaluate the effect of $\tau_c$ on utility and robustness in Section~\ref{sec:ev:correl}. We choose the data size as 1,000 to show the robustness of the proposed scheme for a relatively small data. As we show via experiments later, robustness increases with the increasing data size because as data size increases, we obtain more fingerprinted data points for the same fingerprinting probability $p$. As expected, utility of the data owner (as introduced in Section~\ref{sec:system}) decreases linearly with increasing fingerprinting probability and we observed that the average utility is $0.8$ when $p$ is $0.1$.

We expect to change the value of approximately $p\cdot l = 100$ data points as fingerprint for each SP, and (as discussed in Section~\ref{sec:use-boneh-shaw}) we used approximately half of the fingerprinted data points for Boneh-Shaw codes. We set the number of Boneh-Shaw codes ($c$) as $10$ and the block size of Boneh-Shaw codes ($r$) as $5$. Hence, $(c-1)\cdot r = 45$ fingerprinted data points of first SP were used for Boneh-Shaw codes. 
Another design parameter in the algorithm is $\theta$, which is used to dynamically adjust fingerprinting probability $p$ to keep the number of fingerprinted data points close to $p \cdot l$. $\theta$ can have any value in the range $\lbrack0,1)$. Although the selection of $\theta$ does not have a significant effect on the average number of fingerprinted data points in each shared copy, it has a significant effect on the standard deviation. As shown in Table~\ref{table:theta}, increasing $\theta$ decreases the standard deviation and the number of fingerprinted data points for each SP is close to each other for higher $\theta$ values. Therefore, $\theta$ values close to 1 provide almost $p\cdot l = 100$ fingerprinted data points for each SP. However, for $\theta$ values close to 1, the fingerprinting probability becomes close to $0$ for some blocks within the algorithm where $p \cdot(1 - \theta)$ is used as fingerprinting probability. This may cause an attacker to understand the values of the data points in these blocks. For instance, if the algorithm adds two fingerprints in the first block (including $1/p = 10$ data points), the fingerprinting probability reduces to $p \cdot(1 - \theta)$ (close to $0$ for higher values of $\theta$) in the second block. Therefore, an attacker may understand that all data values in the second block are the original values with very high probability by detecting two fingerprints in the first block (e.g., via a correlation or collusion attack). Hence, we set $\theta = 0.5$ in our experiments, which provides a reasonable standard deviation, as shown in Table~\ref{table:theta}.
We repeated all experiments 10 times for each individual in the dataset (totally 99 individuals), and hence all results are given as the average of 990 executions. 

\begin{table}[ht]
\small
\centering
\caption{The effect of $\theta$ on the standard deviation of the number of fingerprinted data points in each fingerprinted copy.}
\vspace{-2mm}
\begin{tabular}{ c c c c c  }
\hline
$\theta$ & 0 & 0.25 & 0.5 & 0.75  \\
\hline
Standard Deviation & 10.17 & 4.20 & 2.31 & 1.68 \\
\hline
\end{tabular}
\vspace{-4mm}
\label{table:theta}
\end{table} 

\subsection{Flipping and Subset Attacks}
\label{sec:ev:flip}

\begin{figure}[ht]
\centering
\includegraphics[width=9cm,keepaspectratio]{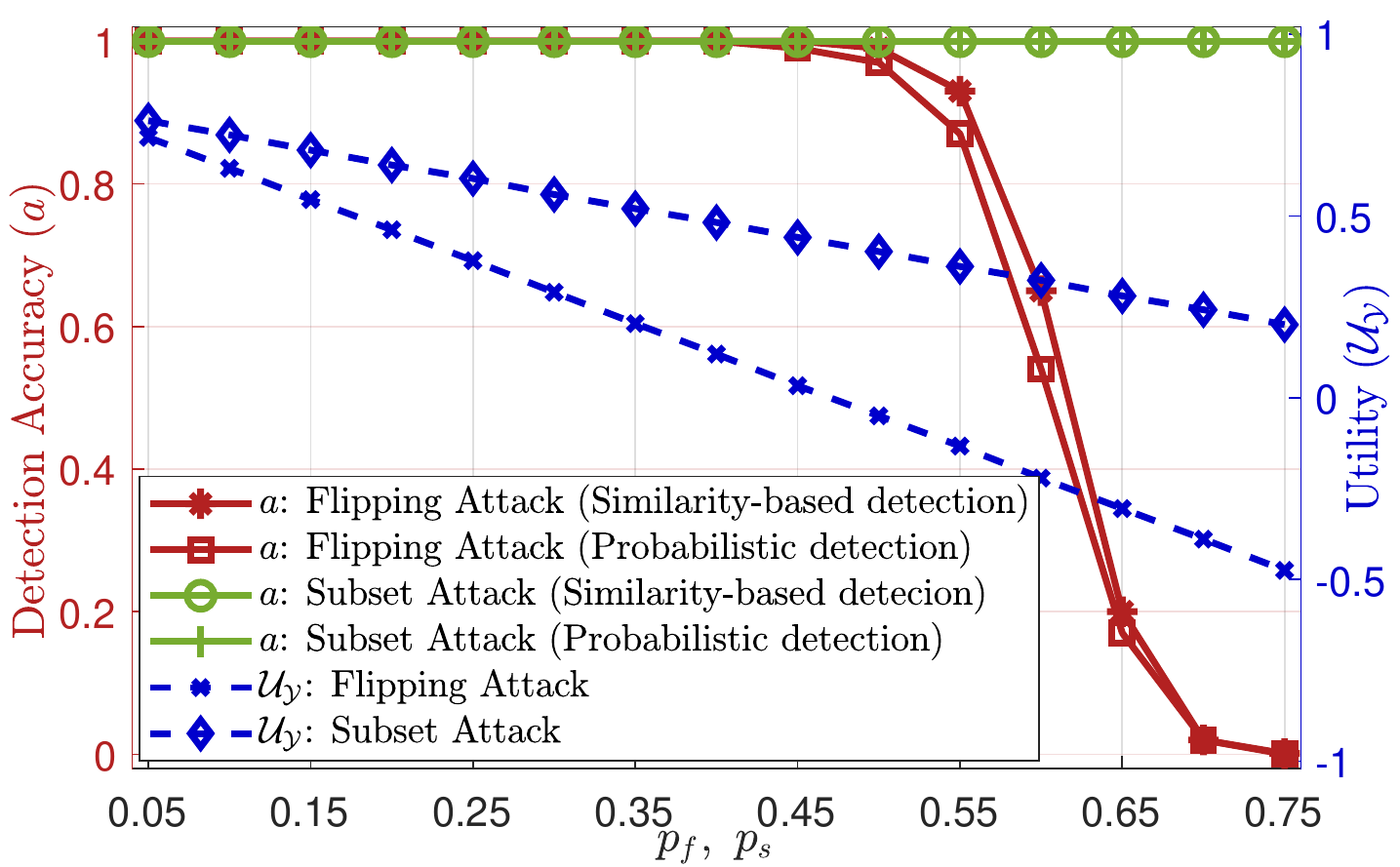}
\caption{Fingerprint robustness against flipping and subset attacks using similarity-based detection and probabilistic detection for different values of flipping probability ($p_f$) in the flipping attack and removing probability ($p_s$) in the subset attack. 
Right $y$-axis is used to show the utility of the attacker ($\mathcal{U}_\mathcal{Y}$, as defined in Section~\ref{sec:attack}) after flipping and subset attacks.}
\label{fig:flipsubset}
\end{figure}

In this experiment, we compared the flipping and subset attacks in terms of their effect on the fingerprint robustness. Also, to compare the similarity-based and probabilistic detection techniques (which are the basic building blocks of the proposed detection algorithm in Section~\ref{sec:collusion:detection}), we implemented them for this experiment. We set the total number of SPs (that received the data owner's fingerprinted data) to 1,000. Figure~\ref{fig:flipsubset} shows the accuracy ($a$) of the similarity-based detection technique and probabilistic detection technique for different values of $p_f$ (probability of flipping a data point in flipping attack) and $p_s$ (probability of removing a data point in subset attack). From these results, we can conclude that (i) similarity-based detection provides slightly better accuracy than the probabilistic detection, (ii) flipping attack is more powerful than subset attack, and (iii) the attacker needs to flip at least half of the data points to avoid being detected, which decreases the utility of the attacker $\mathcal{U}_\mathcal{Y}$ (defined in Section~\ref{sec:attack}) to negative values. For the rest of the experiments, we use the detection algorithm described in Section~\ref{sec:collusion:detection}, which is based on the similarity-based detection technique (that is shown to perform better than probabilistic detection).

To observe the effect of the total number of SPs on robustness, we computed $a$ by changing the number of SPs that receives a fingerprinted copy from Alice and we show our results in Table~\ref{table:NumSP}. We observed similar results with Figure~\ref{fig:flipsubset} even the data is shared with 10,000 SPs.
If data is shared with 10,000 SPs and any of these SPs leaks its copy by flipping up to $44\%$ of the data points, similarity-based technique detects the guilty SP with $99\%$ accuracy. As we also show in Figure~\ref{fig:flipsubset}, a malicious SP obtains almost 0 utility by flipping $44\%$ of the data points. We computed the utility of the attacker ($\mathcal{U}_\mathcal{Y}$) as described in Section~\ref{sec:attack} by assigning equal utility for each data point. Hence, an attacker needs to share data with negative utility after subset or flipping attack to avoid being detected.

\begin{table}[ht]
\small
\centering
\caption{The effect of the number of SPs on the fingerprint robustness. The table shows the amount of flipping ($p_f$) needed by the attacker to decrease the robustness below $99\%$ and $90\%$. For instance, if data is shared with 10,000 SPs, the attacker (one of these SPs) needs to flip at least $49\%$ of its data points to decrease the robustness of the data owner below $90\%$.}
\vspace{-3mm}
\begin{tabular}{ c c c c c  }
\hline
& \multicolumn{4}{c}{The number of SPs} \\
\hline
 & 10 & 100 & 1,000 & 10,000  \\
\hline
$a \leq 99\%$ & 0.52 & 0.50 & 0.46 & 0.44 \\
$a \leq 90\%$ & 0.57 & 0.54 & 0.51 & 0.49 \\
\hline
\end{tabular}
\vspace{3mm}
\label{table:NumSP}
\end{table}

\subsection{Correlation Attack}
\label{sec:ev:correl}

\begin{figure}[ht]
\centering
\includegraphics[width=9cm,keepaspectratio]{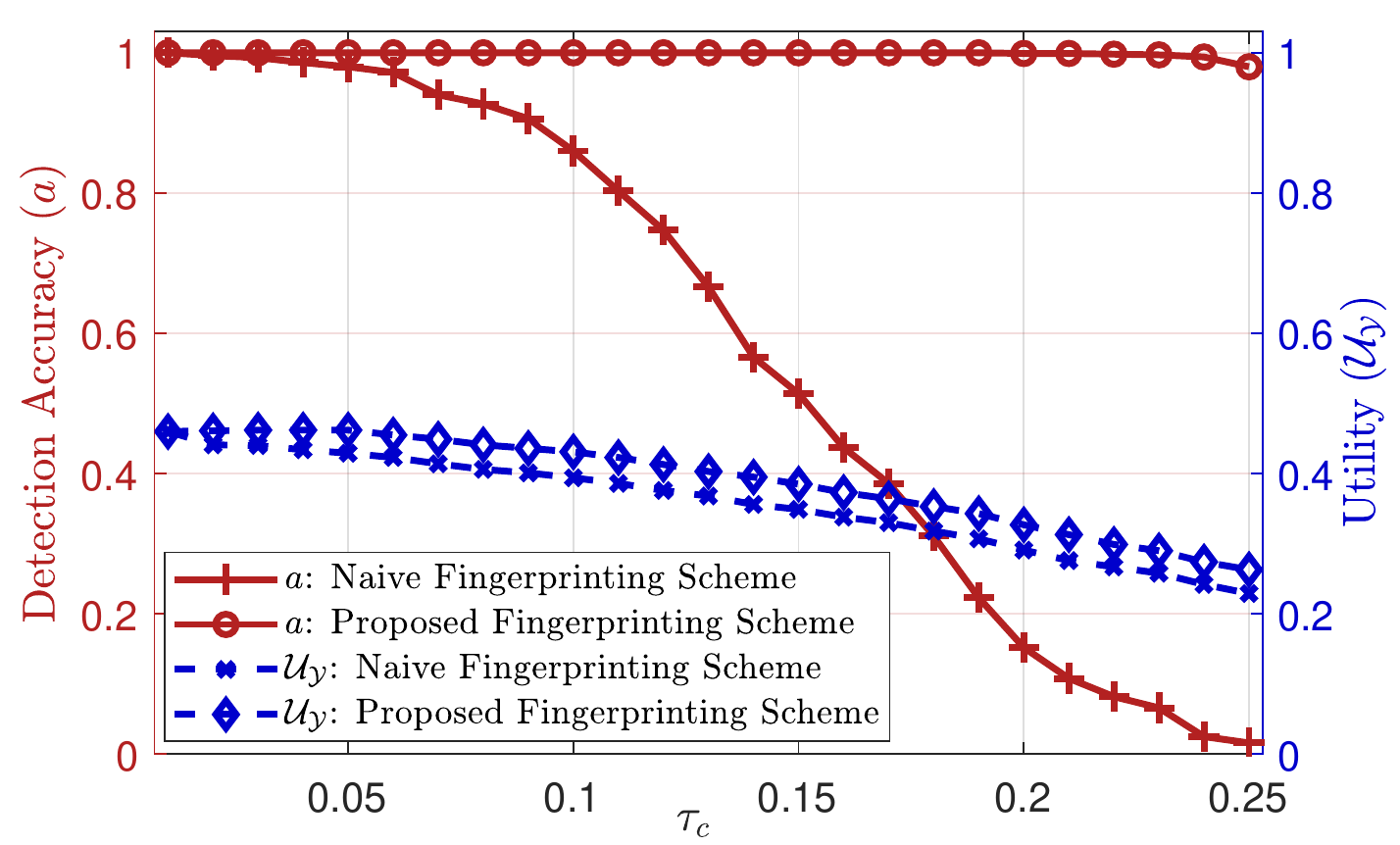}
  \vspace{-4mm}
\caption{Fingerprint robustness of the proposed algorithm and the naive algorithm (in Section~\ref{sec:scheme}) 
against correlation attack for different values of correlation attack threshold. $p_f$ was selected as $0.2$. Right $y$-axis is used to show the utility of the attacker ($\mathcal{U}_\mathcal{Y}$) after correlation attack.}
\label{fig:robCor}
\end{figure}

We implemented the correlation attack described in Section~\ref{sec:threat:correlation} to evaluate the robustness of the proposed scheme. We set the total number of SPs (that received the data owner's fingerprinted data) to 1,000. As before, we set the correlation threshold of the algorithm (decided by data owner) as $\tau = 0.05$. Therefore, the fingerprinted copies did not include consecutive pairs of data points whose correlation is less than 0.05. Note however that the correlation threshold of the attack $\tau_c$ is determined by the attacker. We also implemented the naive fingerprinting scheme described in Section~\ref{sec:scheme}, in which each data point is fingerprinted with probability $p$. As described in Section~\ref{sec:threat:correlation}, in correlation attack, data points whose correlation with the previous data point is less than the correlation attack threshold $\tau_c$ is flipped and the remaining data points are flipped with probability $p_f$. Figure~\ref{fig:robCor} shows the comparison of the proposed scheme with the naive approach for different values of $\tau_c$ when $p_f = 0.2$. The proposed scheme provides $100\%$ detection accuracy up to $\tau_c = 0.2$ and accuracy decreases to $98\%$ when $\tau_c = 0.25$. However, as also shown in Figure~\ref{fig:robCor}, the utility of the attacker ($\mathcal{U}_\mathcal{Y}$) reduces to $0.263$ when $\tau_c = 0.25$. The utility of the attacker is almost constant for $\tau_c < 0.05$, because the proposed algorithm guarantees that there is no pairs of data points whose correlation is less than correlation threshold of the algorithm ($\tau = 0.05$). We also observed that the naive approach is not robust against correlation attacks and the attacker can easily prevent detection by utilizing the correlations in the data. This clearly shows the importance of considering correlations in the data within the fingerprinting algorithm.

Furthermore, we observed the robustness against correlation attack for different values of fingerprinting probability $p$. As shown in Table~\ref{table:pCorrel}, the robustness increases when the $p$ increases because the number of fingerprinted data points is directly proportional to $p$. We also obtained similar results (as in Table~\ref{table:NumSP}) for the effect of the total number of SPs on robustness.

\begin{table}[ht]
\small
\centering
\caption{The effect of $p$ on the detection accuracy against correlation attack.}
\begin{tabular}{ c c c c c c c c c c c }
\hline
 $p$ & 0.02 & 0.04 & 0.06 & 0.08 & 0.1 & 0.12 & 0.14 & 0.16 & 0.18 \\
\hline
$a$ & 0.24 & 0.6 & 0.88 & 0.94 & 0.98 & 0.99 & 0.99 & 1 & 1 \\
\hline
\end{tabular}
\label{table:pCorrel}
\end{table}

\subsection{Collusion Attack}
\label{sec:ev:collusion}

\begin{figure}[ht]
\centering
\includegraphics[width=9cm,keepaspectratio]{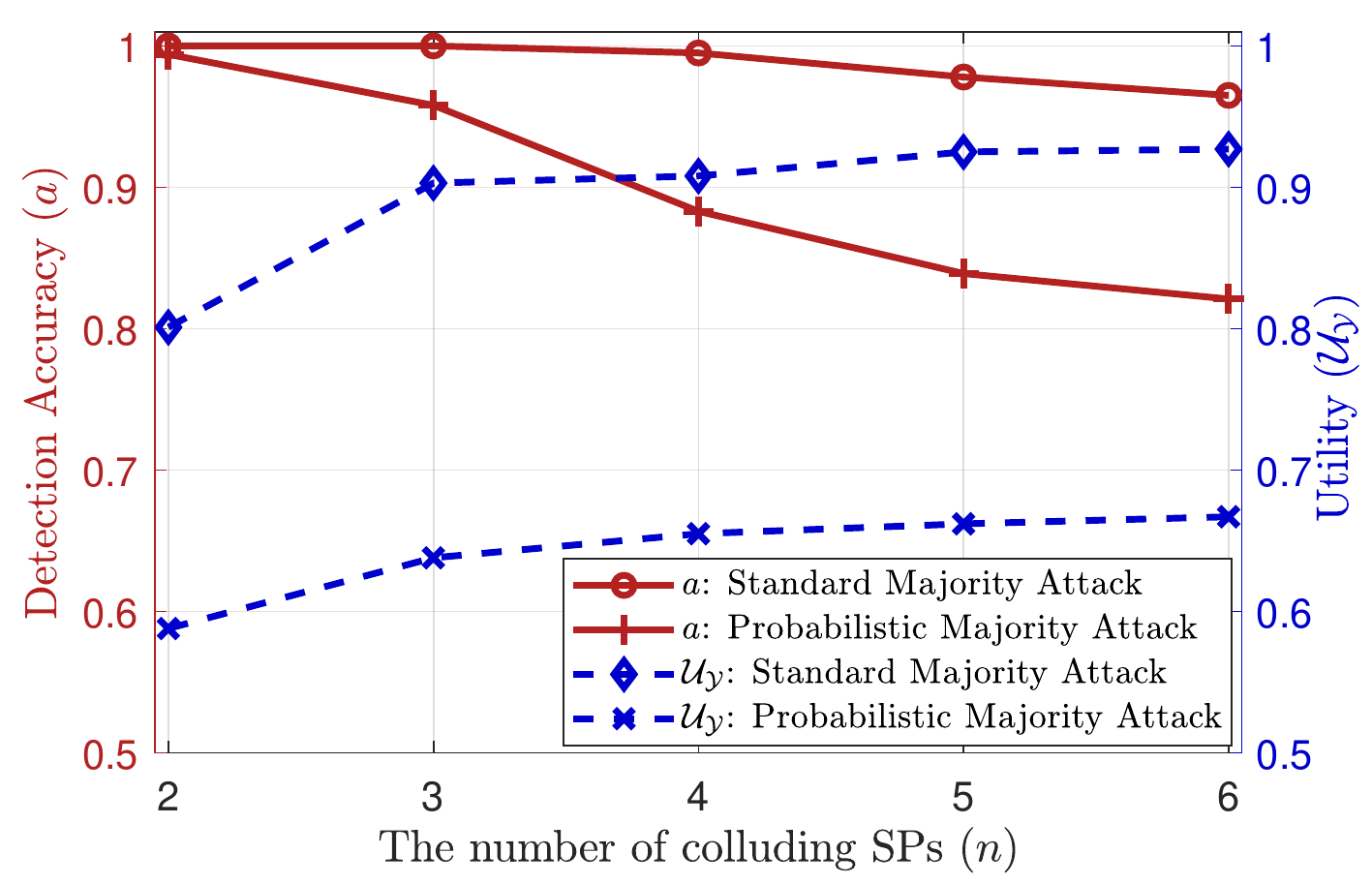}
\caption{Fingerprint robustness against standard and probabilistic majority attacks for different values of $n$ (number of colluding SPs). Right $y$-axis is used to show the utility of the attackers ($\mathcal{U}_\mathcal{Y}$) after performing majority attack.}
\label{fig:colMajority}
\end{figure}

As mentioned in Section~\ref{sec:threat:collusion}, in a standard collusion attack, colluding SPs select (and share) the most commonly observed value for each data point. In this attack, although the utility of the data leaked by the colluding SPs ($\mathcal{U}_\mathcal{Y}$) is high, detection techniques can detect the colluding SPs with high probability. Hence, we defined a probabilistic majority attack in Section~\ref{sec:collusionattack} that also includes correlation and flipping attacks. With this attack, the colluding SPs decrease the accuracy of the detection algorithm by reducing the utility of data. In this section, we first compare the utility and robustness under these two attacks. We set $c = 10$ and $r = 5$ as the Boneh-Shaw parameters. Hence, we can create 10 different Boneh-Shaw codewords with block size of $5$. Note that $c$ is the number of codewords and Alice can share her data more than $c$ SPs in the proposed scheme by repeating codewords as we discussed before. We set the total number of SPs (that received data owner's data) to 10, $\tau_c = 0.1$, and $p_f = 0.1$. We quantified both utility and fingerprint robustness for different values of number of colluding SPs ($n$). As shown in Figure~\ref{fig:colMajority}, using probabilistic majority attack decreases the colluding SPs' probability of being detected by reducing the data utility. Since the probabilistic majority attack is more powerful than the standard one for the colluding SPs, we perform the probabilistic majority attack for the rest of the experiments.

As mentioned, Boneh-Shaw codes are not designed to be robust against majority attacks. They assume the colluding SPs decide the value of a data point randomly if they observe more than one value in their copies. Hence, Boneh-Shaw codes do not provide guarantees against the proposed probabilistic majority attack (including flipping). Furthermore, in order to provide the robustness guarantees of Boneh-Shaw codes, the number of fingerprinted points needs to be high. For instance, to create Boneh-Shaw codes for $c = 10$ with $90\%$ robustness guarantee, the block size needs to be greater than 1,000 and we need more than 10,000 fingerprinted data points. Fingerprinting such a high number of data points also decreases the utility of the data owner ($\mathcal{U}_i$) significantly. Therefore, direct use of Boneh-Shaw codes do not provide guarantees, especially for small data sizes (we only use approximately $p \cdot l = 100$ data points for fingerprinting in our experiments). To show these shortcomings, we implemented Boneh-Shaw codes as a standalone fingerprinting scheme and observed its robustness against the probabilistic majority attack using the same parameters with the previous experiment (Figure \ref{fig:colMajority}). We observed $55\%$ detection accuracy when the number of colluding SPs was 2 (for the same scenario, the accuracy of the proposed scheme is more than $99\%$, as shown in Figure~\ref{fig:colMajority}). Therefore, we conclude that using Boneh-Shaw codes as a standalone fingerprinting scheme does not provide robustness against probabilistic majority attack (which includes correlation and flipping attacks). While our proposed scheme utilizes the Boneh-Shaw codes to increase its robustness against collusion attacks, Algorithm~\ref{alg:alg2} generates unique fingerprints to provide robustness against correlation and flipping attacks.

We also evaluated the robustness of the proposed scheme by increasing the number of SPs receiving fingerprinted copies Alice's (data owner's) data. Due to the size of the data, we selected $c = 10$ and $r = 5$ in our experiments. We can create $10$ different Boneh-Shaw codewords with these parameters. In such a scenario, if Alice wants to share her data with more than $10$ SPs, she assigns the same $10$ codewords to other SPs. Therefore, Boneh-Shaw codewords of SPs whose index values are equivalent in modulo $c = 10$ are same where their remaining unique fingerprints are generated by the Algorithm~\ref{alg:alg2}. For instance, all $SP_1$, $SP_{11}$, and $SP_{21}$ receive the same Boneh-Shaw fingerprints in our proposed scheme. We show the robustness of the proposed scheme for different values of number of SPs in Table~\ref{table:colNumSP}. We observed that the detection accuracy of Alice reduces when she shares her data with more SPs. For instance, increasing the number of SPs from 50 to 100 decreased the accuracy from 0.824 to 0.759 in our experiments when the number of colluding SPs ($n$) was selected as 3. Note that flipping probability ($p_f$) was selected as $0.1$ in these experiments.  

\begin{table}
\small
\centering
\caption{Fingerprint robustness ($a$) of the proposed scheme for different values of number of SPs (that received data owner's data) ($|\mathcal{S}|$) and $n$ (the number of colluding SPs).}
\begin{tabular}{ c c c c c c }
\hline
& \multicolumn{5}{c}{Number of SPs} \\
\hline
 & 50 & 100 & 150 & 200 & 250 \\
\hline
$n = 2$ & 0.995 & 0.992 & 0.99 & 0.989 & 0.981 \\
\hline
$n = 3$ & 0.824 & 0.759 & 0.69 & 0.667 & 0.614 \\
\hline
$n = 4$ & 0.606 & 0.504 & 0.411 & 0.344 & 0.343 \\
\hline
\end{tabular}
\label{table:colNumSP}
\end{table}

\begin{figure}[ht]
\centering
\includegraphics[width=9cm,keepaspectratio]{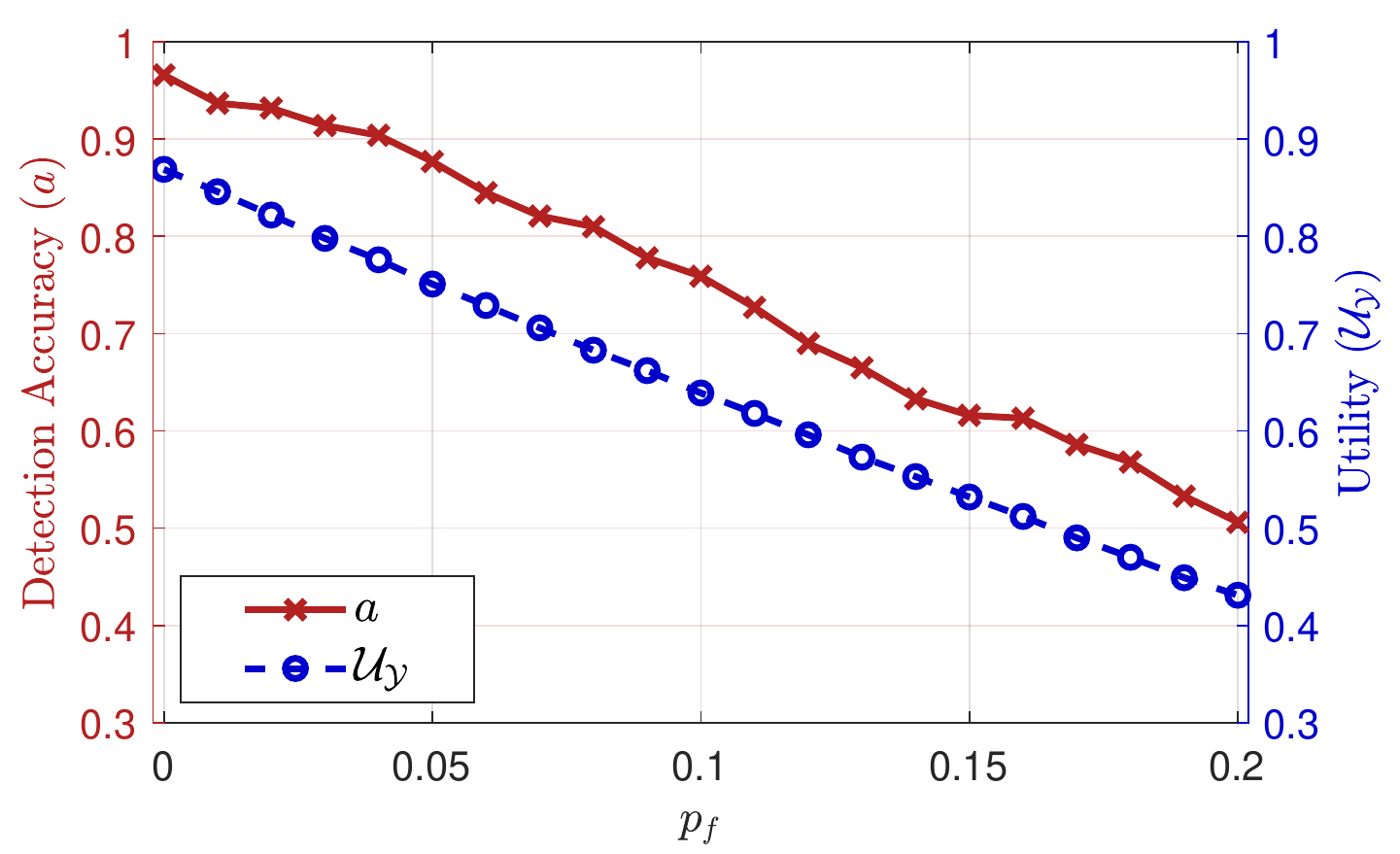}
\caption{Fingerprint robustness and utility of the proposed scheme when probabilistic majority attack is performed using different values of $p_f$ (flipping probability). The number of SPs (that received data owner's data) is 100 and the number of colluding SPs ($n$) is 3.}
\label{fig:colflipprob}
\end{figure}

Next, we evaluated the effect of flipping probability ($p_f$) on fingerprint robustness during a collusion attack. Our results are shown in Figure~\ref{fig:colflipprob}. Increasing $p_f$ reduces the robustness and the utility considerably. When there is no flipping ($p_f = 0$) in the probabilistic majority attack, Alice detects one of $3$ colluding SPs among 100 SPs (that she shared her data) with $97\%$ accuracy. The accuracy becomes less than $90\%$ when $p_f$ is selected as $0.05$, where the utility of the attackers also decreases from 0.87 to 0.75. 

Another important parameter for the fingerprinting scheme is data size. In our experiments we used a data with 1,000 SNPs ($l = 1,000$). Therefore, we just changed the state of approximately 100 data points as a fingerprint when fingerprinting probability $p$ was selected as 0.1. By keeping the same fingerprinting probability, increasing data size allows to change more data points as fingerprint. With more fingerprinted data points, the data owner can detect the colluding SPs with higher accuracy. To show the effect of data size (i.e., $l$) on robustness, we conducted experiments by increasing data size (we kept $c = 10$ and increased $r$ proportional to data size). As shown in Table~\ref{table:colDataSize}, Alice detects one of 3 colluding SPs among 100 SPs with $99.7\%$ accuracy when $l = 5,000$ and $p_f = 0.1$. Thus, we observed that the robustness of the proposed scheme significantly improves with increasing data size.

\begin{table}[ht]
\small
\centering
\caption{Fingerprint robustness ($a$) of the proposed scheme for different $l$ (data size) values. The number of SPs (that received data owner's data) is 100 and the number of colluding SPs ($n$) is 3. Flipping probability in the attack is 0.1.}
\begin{tabular}{ c c c c c c }
\hline
$l$ & 1,000 & 2,000 & 3,000 & 4,000 & 5,000 \\
\hline
$a$ & 0.759 & 0.914 & 0.973 & 0.993 & 0.997   \\
\hline
\end{tabular}
\label{table:colDataSize}
\end{table}

\subsection{Privacy-Preserving Fingerprinting}
\label{sec:ev:privacy}

As we explained in Section~\ref{sec:privacy}, both the proposed scheme and LDP-based mechanisms change the values of some points in data before sharing. However, while the goal of fingerprinting schemes is to provide robustness, LDP-based mechanisms aim to guarantee privacy of the individuals. To show the trade-off between fingerprint robustness and privacy, we first implemented the randomized response (RR) mechanism~\cite{warner1965randomized} which satisfies $\epsilon$-LDP when the state of each data point is correctly shared with probability $e^\epsilon/(e^\epsilon + 2)$ for genomic data with three possible states. Also, each of the incorrect two values can be shared with a probability of $1/(e^\epsilon + 2)$. To compare LDP with the proposed scheme, we set $e^\epsilon/(e^\epsilon + 2) = 0.9$, for which RR satisfies LDP with $\epsilon = 2.89$. As the privacy metric, we used the average estimation error, which is a commonly used to quantify genomic privacy~\cite{wagner2017evaluating}. This metric quantifies the average distance of the copy created by colluding SPs ($\mathcal{Y}$) from the original data ($\mathcal{X}$) as $E = (\sum_{j=1}^{l} ||x_j-y_j||)/l$.

We set the number of SPs (that received the data) to $10$ and performed the probabilistic majority attack described in Section~\ref{sec:collusionattack} with $3$ colluding malicious SPs by setting both of the flipping probability ($p_f$) and the correlation threshold of the colluding SPs ($\tau_c $) to $0.1$. Although RR mechanism is similar to naive probabilistic scheme described in Section~\ref{sec:scheme}, all SPs receive the same copy in RR to guarantee LDP (as discussed in Section~\ref{sec:privacy}). 
Since fingerprint robustness is defined as detecting one colluding SP (in Section~\ref{sec:robustness}), and all SPs have the same copy, we observed $0.3$ detection accuracy ($a$) for the RR mechanism, which is equivalent to randomly accusing any SP that received the data. 
We also observed $E$ as $0.279$ when the probabilistic majority attack was performed for the RR mechanism. 

\begin{figure}
\centering
\includegraphics[width=9cm,keepaspectratio]{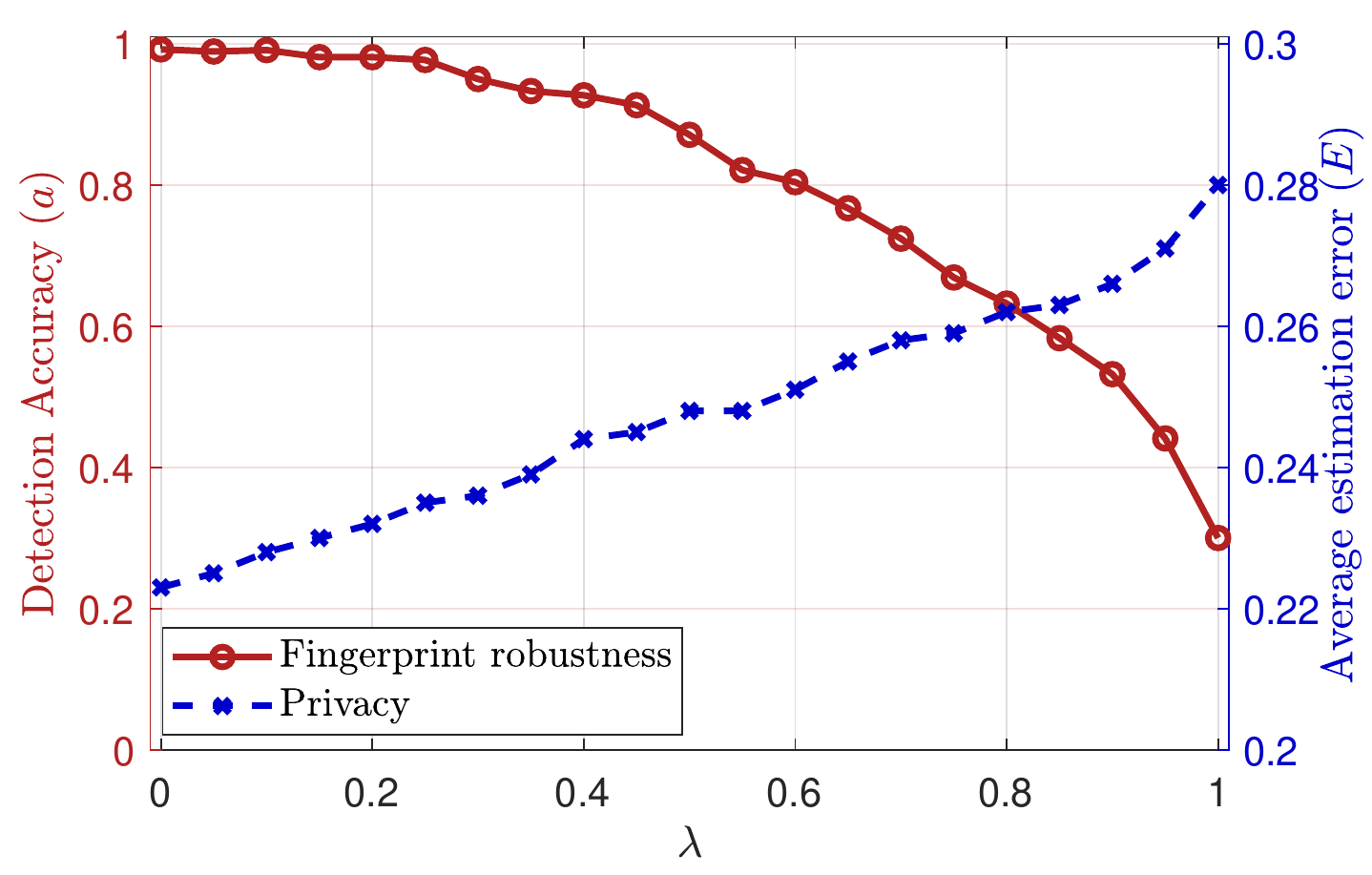}
  \vspace{-2mm}
\caption{Fingerprint robustness and privacy of the hybrid approach for different values of $\lambda$. Probabilistic majority attack was performed by $3$ colluding malicious SPs among 10 SPs that received the data. Both $p_f$ and $\tau_c$ were chosen as 0.1.}
  \vspace{-2mm}
\label{fig:priRob}
\end{figure}

Next, we also implemented the hybrid scheme described in Section~\ref{sec:privacy}. In this scheme, the parameter $\lambda$ determines the amount of overlapping fingerprints which are included in the copy of each SP. Hence, the proposed fingerprinting scheme (in Section~\ref{sec:use-boneh-shaw}) is equivalent to the hybrid approach when $\lambda=0$. Figure~\ref{fig:priRob} shows both fingerprint robustness ($a$) and privacy ($E$) provided by the hybrid scheme for different values of $\lambda$. When $\lambda$ is selected as $1$ we observed similar robustness compared to the RR mechanism. However, the hybrid scheme provides slightly higher error (better privacy) than the RR mechanism since the correlations in the data are considered in hybrid scheme as opposed to the RR mechanism. 
Privacy improves with decreasing fingerprint robustness and when $\lambda$ increases, the loss in the robustness is significantly higher than the gain in the privacy. We observed that for values of $\lambda$ that are below $0.5$, the proposed hybrid scheme provides both high fingerprint robustness and reasonable privacy.  

%% file: sections/discussion.tex
\section{Discussion}\label{sec:discuss}

In this section, based on our experimental results we further discuss about considering privacy and liability together, the practicality of the proposed scheme, and the application of the proposed scheme to different domains.

\subsection{Fingerprint Robustness and Privacy}

When individuals share their personal data with SPs, they want to protect their privacy as well as to identify
source of a potential data leakage in case of illegal distribution of their data. As shown in Section~\ref{sec:ev:privacy}, techniques to protect the privacy of individuals (such as LDP-based mechanisms) do not let individuals to detect the source of unauthorised sharing. Similarly, fingerprinting techniques to provide liability for unauthorized sharing of personal data do not consider the privacy of the individuals. In this work, for the first time, we studied liability and privacy together and proposed a hybrid approach to provide privacy and fingerprint robustness together. Although it is not possible to achieve the highest privacy and the highest fingerprint robustness at the same time (due to their conflicting objectives), we showed that it is possible to achieve a scheme that provides reasonable privacy and fingerprint robustness at the same time. We anticipate that this work will pave the way towards a new research direction to achieve privacy and liability using a single mechanism. We plan to study this problem extensively in the future and develop novel mechanisms that provide formal privacy and robustness guarantees.

\subsection{Complexity and Practicality}

In the proposed fingerprinting algorithm (Algorithm~\ref{alg:alg2}), each data point sequentially decides on a probability for each possible value in set $\mathcal{D}$ and inserts the fingerprints accordingly. Hence, the complexity of fingerprinting algorithm is $\Theta(l \cdot m)$. To detect the guilty SP in case of data leakage, the data owner needs to compare all fingerprint patterns (given to all SPs) with the leaked data. Since the expected value of fingerprinted data points is $p \cdot l$ in each fingerprinted copy, the complexity of the detection algorithm is $\Theta(|\mathcal{S}| \cdot p \cdot l)$, where $|\mathcal{S}|$ is the number of SPs that received a fingerprinted copy. Note that this is also the storage complexity for Alice if she stores all the fingerprint patterns. As mentioned before, if Alice does not want to store all fingerprint patterns, she can just store the seed value for each SP and check the similarity of fingerprint patters by running the fingerprinting algorithm again in case of data leakage. Note that this slightly increases the complexity of detection algorithm since it requires Alice to run the fingerprinting algorithm along with the detection algorithm. Thus, we conclude that the running times of both fingerprinting algorithm and detection algorithm grow linearly with the design parameters.

To show the practicality of the proposed scheme, we observed its running time using a computer with 1.8 GHz Dual-Core Intel Core i5 processor and 8 GB memory. Based on our experiments, we measured the average running time of fingerprinting algorithm to create one fingerprinted copy as $0.15$ ms. and the average running time of detection algorithm as $3.11$ ms. when $|\mathcal{S}| = 1000$, $l = 1000$, $p = 0.1$, and $m = 3$. These results also show the efficiency and practicality of the proposed scheme.

\subsection{Application of the Proposed Scheme to Other Domains}
\label{sec:application}

In the evaluations, we implemented the proposed fingerprinting scheme for genomic data sharing. However, the proposed scheme can be used for sharing other types of personal data, such as location data. 
Here, we briefly discuss the differences for the application of the proposed algorithm between genomic data and location data.
For different data types, the main difference from genomic data will be the set of possible values for data points ($\mathcal{D}$). For genomic data, $\mathcal{D}$ contains three values for all data points. However, when location data is considered, $\mathcal{D}$ can be the set of point of interests (POIs) at which Alice can be located at a specific time. Hence, for each data point, the set $\mathcal{D}$ may be different. The size of $\mathcal{D}$ for location data will also be larger, and hence the running time of the fingerprinting algorithm will be higher for location data. In terms of robustness, we expect an improvement for location data application due to having more possible values for each data point. In genomic data, when a data point is fingerprinted, its value must be one of two remaining values, and hence providing the same fingerprint for a data point to multiple SPs is very likely for genomic data. However, when Algorithm~\ref{alg:alg2} adds a fingerprint to a location data point for two different SPs, the probability of adding the same fingerprint is lower (due to size of $\mathcal{D}$), which improves the uniqueness of the fingerprints. As discussed before, uniqueness of the fingerprints increases fingerprint robustness. 

%% file: sections/conclusion.tex
\vspace{1mm}
\section{Conclusion}\label{sec:conc}

We have proposed a probabilistic fingerprinting scheme that also considers the correlations in the data during fingerprint insertion. 
First, we have shown how to assign probabilities for the sharing decision of each data point that are consistent with the inherent correlations in the data. 
Then, we have described the integration of Boneh-Shaw codes into the proposed algorithm to improve fingerprint robustness against collusion attacks. 
We have also proposed a detection algorithm that initially selects the suspects based on similarity scores and decides the guilty SP using the detection technique of Boneh-Shaw codes. 
Furthermore, to provide privacy along with fingerprint robustness, we have proposed a hybrid approach that controls the trade-off between privacy and fingerprint robustness. 
Our experimental results on genomic data show that the proposed fingerprinting scheme is robust against a wide range of attacks and it can also provide reasonable privacy by slightly degrading the fingerprint robustness. The proposed scheme is a first step for sharing personal data with service providers by providing both liability and privacy guarantees at the same time.